\pdfoutput=1

\documentclass[11pt]{article}

\usepackage{ielabarxiv}
\usepackage[round,authoryear]{natbib}
\usepackage{latexsym}
\usepackage[T1]{fontenc}
\usepackage[utf8]{inputenc}
\usepackage{microtype}
\usepackage{inconsolata}
\usepackage{array}
\usepackage{fancyvrb}
\usepackage{float}
\usepackage{placeins}
\usepackage{pgfplots}
\pgfplotsset{compat=1.18}
\usetikzlibrary{arrows.meta,patterns,patterns.meta}

\definecolor{cbGray}{HTML}{7F7F7F}
\definecolor{cbBlue}{HTML}{0072B2}
\definecolor{cbVermillion}{HTML}{D55E00}
\definecolor{cbOrange}{HTML}{E69F00}
\definecolor{cbBluishGreen}{HTML}{009E73}
\definecolor{seqDark}{HTML}{08519C}
\definecolor{seqMed}{HTML}{4292C6}
\definecolor{seqLight}{HTML}{C6DBEF}

\preprintheader{arXiv preprint.}

\panelTitle{Search, Inspect, Fetch:\\
Exploiting Structure-Aware Boolean Retrieval for Deep-Search Agents}
\panelAuthors{%
Shuai Wang,\quad Haodong Chen,\quad Yu Yin,\quad Shengyao Zhuang\par
\vspace{0.2em}
Bevan Koopman,\quad Guido Zuccon}
\panelAffiliation{%
$^{1}$The University of Queensland, Brisbane, Australia\quad
$^{2}$CSIRO, Brisbane, Australia}
\panelEmails{%
\{shuai.wang2, y.yin1, s.zhuang, g.zuccon\}@uq.edu.au\\
haodong.chen1@student.uq.edu.au\quad bevan.koopman@csiro.au}
\panelLinks{%
\projectlink{https://ielab.io/skim-search-agent/}\\
\githublink{https://github.com/ielab/skim-search-agent}\\
\hflink{https://huggingface.co/collections/wshuai190/sieve}}

\hypersetup{
  pdftitle={Search, Inspect, Fetch: Exploiting Structure-Aware Boolean Retrieval for Deep-Search Agents},
  pdfauthor={Shuai Wang, Haodong Chen, Yu Yin, Shengyao Zhuang, Bevan Koopman, Guido Zuccon}
}

\begin{document}
\begin{titlepanel}
Existing deep-search agents use a Search--Visit workflow that retrieves whole webpages without
considering the structure they expose through titles, headings, sections, and metadata. This
prevents agents from directly constraining retrieval to parts of a webpage and often carries
irrelevant content into their context. We introduce \textsc{Sieve}, a search--inspect--fetch
strategy driven by a Boolean Query Language (BQL): it searches webpage fields to filter candidates,
uses an interchangeable ranker to order them, presents structure-rich result cards for inspection,
and fetches only selected
sections. Across three QA collections, \textsc{Sieve} is more accurate than the strongest
conventional Search--Visit configuration on each collection while using $20.7$--$50.6\%$ fewer
tokens. Boolean filtering improves every tested ranker, and the accuracy--context advantage persists
across retriever choices and agent backbones. Our implementation is included in the
SkimSearchAgent library.

\panelLogos{\logoielab\logosep\logouq}
\end{titlepanel}

\setcounter{figure}{1}
\begin{figure}[H]
\centering
\includegraphics[width=\textwidth]{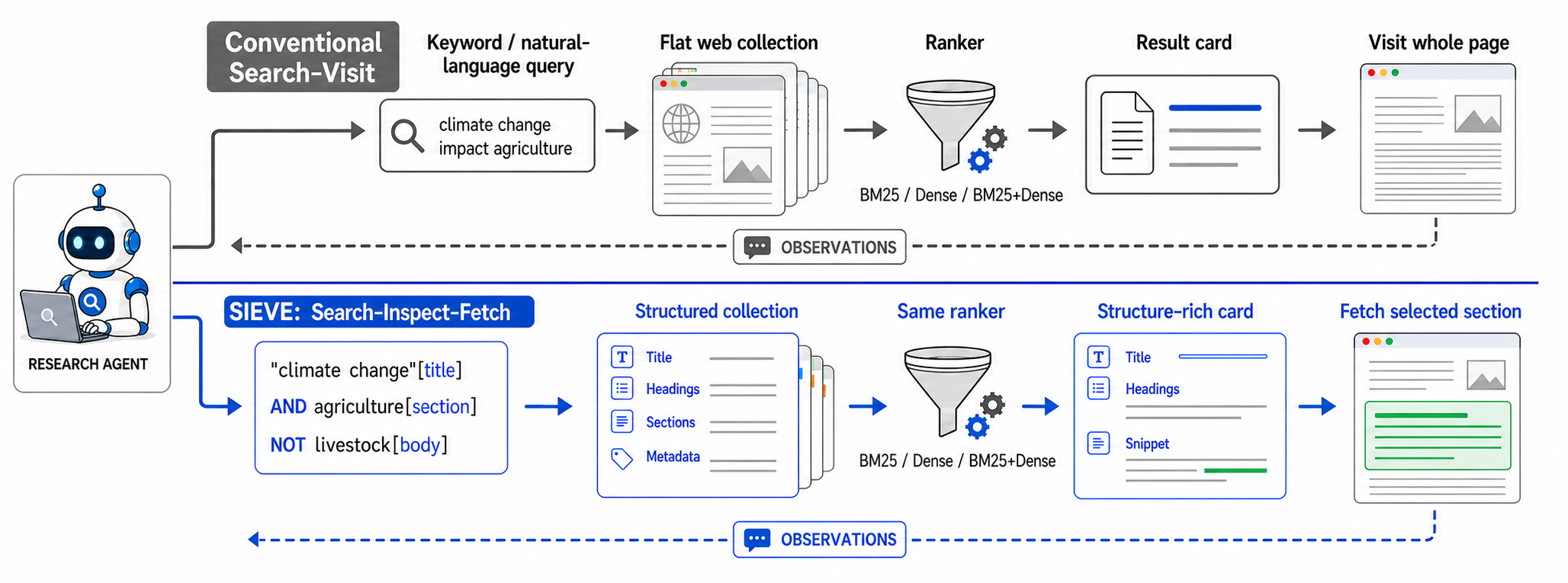}
\caption{Workflow comparison. Search--Visit returns whole pages; \textsc{Sieve} selects with BQL,
shows structured result cards, and fetches named sections. Both can use the same agent and ranker.}
\label{fig:architecture}
\end{figure}

\clearpage

\section{Introduction}
\label{sec:introduction}

Deep-search agents answer difficult questions through an iterative Search--Visit loop. At each
iteration, an agent formulates a subquery, inspects a ranked list of short webpage snippets, visits
selected webpages, and uses the resulting evidence either to answer or to formulate the next query.
Unlike single-pass retrieval, this Search--Visit loop lets the search evolve as evidence accumulates
\citep{tongyi2025,li2025webthinker,browsecompplus}. Yet each visit remains whole-webpage: after choosing
a result from its short representation, the agent receives the complete webpage even when only one
section is relevant.

\setcounter{figure}{0}
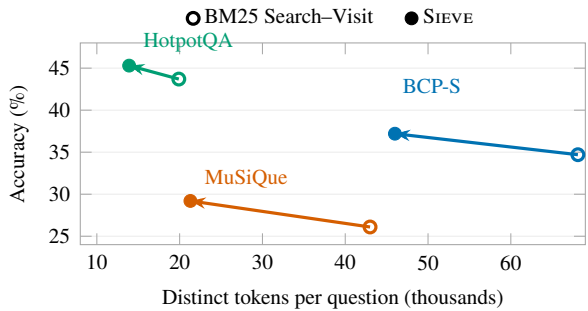
\begin{figure}[H]
\centering
\begin{tikzpicture}
\begin{axis}[
  width=0.5\linewidth,
  height=4.25cm,
  xlabel={Distinct tokens per question (thousands)},
  ylabel={Accuracy (\%)},
  xmin=8,xmax=69,
  ymin=24,ymax=48,
  xtick={10,20,30,40,50,60},
  ytick={25,30,35,40,45},
  tick label style={font=\scriptsize},
  label style={font=\scriptsize},
  axis line style={gray!70},
  tick style={gray!70},
  ymajorgrids,
  grid style={gray!15},
  legend style={
    at={(0.5,1.02)},
    anchor=south,
    legend columns=2,
    draw=none,
    fill=none,
    font=\scriptsize,
    /tikz/every even column/.append style={column sep=10pt}
  },
  clip=false,
]

\addlegendimage{only marks,mark=o,mark size=2.2pt,very thick,color=black,fill=white}
\addlegendentry{BM25 Search--Visit}
\addlegendimage{only marks,mark=*,mark size=2.4pt,color=black}
\addlegendentry{\textsc{Sieve}}

\addplot[-{Stealth[length=2mm]},very thick,color=cbBlue]
  coordinates {(68.1,34.7) (46.0,37.2)};
\addplot[-{Stealth[length=2mm]},very thick,color=cbBluishGreen]
  coordinates {(19.9,43.7) (13.9,45.3)};
\addplot[-{Stealth[length=2mm]},very thick,color=cbVermillion]
  coordinates {(43.0,26.1) (21.3,29.2)};

\addplot[only marks,mark=o,mark size=2.2pt,very thick,color=cbBlue,fill=white]
  coordinates {(68.1,34.7)};
\addplot[only marks,mark=*,mark size=2.4pt,color=cbBlue]
  coordinates {(46.0,37.2)};
\addplot[only marks,mark=o,mark size=2.2pt,very thick,color=cbBluishGreen,fill=white]
  coordinates {(19.9,43.7)};
\addplot[only marks,mark=*,mark size=2.4pt,color=cbBluishGreen]
  coordinates {(13.9,45.3)};
\addplot[only marks,mark=o,mark size=2.2pt,very thick,color=cbVermillion,fill=white]
  coordinates {(43.0,26.1)};
\addplot[only marks,mark=*,mark size=2.4pt,color=cbVermillion]
  coordinates {(21.3,29.2)};

\node[font=\scriptsize,anchor=south west,text=cbBlue]
  at (axis cs:45.7,40.8) {\textsc{BCP-S}};
\node[font=\scriptsize,anchor=south west,text=cbBluishGreen]
  at (axis cs:14.4,45.7) {HotpotQA};
\node[font=\scriptsize,anchor=south west,text=cbVermillion]
  at (axis cs:21.8,29.5) {MuSiQue};
\end{axis}
\end{tikzpicture}
\caption{Higher and farther left is better. Accuracy is the judge verdict on
\textsc{BrowseComp-Plus-Structured} (BCP-S) and exact match on HotpotQA and MuSiQue. Relative to
BM25 Search--Visit, \textsc{Sieve} has higher observed accuracy while using
$30.4$--$50.6\%$ fewer distinct tokens.}
\label{fig:teaser}
\end{figure}

\setcounter{figure}{2}

Whole-webpage visits are a poor match for how webpages are organized. HTML commonly exposes titles,
headings, sections, dates, authors, and other fields that can be parsed directly. Current agent
workflows may display some of these cues, but they do not let the agent use webpage structure to
control both retrieval and content access. We argue that structure should be preserved throughout the
research process. An agent should \emph{search} webpage fields to identify eligible sources,
\emph{inspect} their structure to locate relevant sections, and \emph{fetch} only the content
needed. We call this strategy \emph{search--inspect--fetch}.

Making structure actionable requires retrieval to do more than rank webpages. The agent must also
define which webpages are eligible for inspection. A question may require terms in particular
fields, a date within a specified range, or the absence of an unwanted concept. A ranker can prefer
sources with these properties, but cannot guarantee that every result satisfies them.

Fielded Boolean retrieval was designed for this selection problem. It composes exact constraints
over webpage fields to produce an eligible set, after which a lexical or dense ranker can order
the remaining webpages. Boolean selection therefore determines what may be considered, while
ranking determines what should be inspected first. This division also keeps ranking modular: the
same Boolean-selected set can be ordered by BM25, a dense retriever, or a fusion of both.

Consider an agent looking for government heatwave plans published after 2020 that recommend school
closures but are not workplace-safety guidance. During search, it can require ``heatwave plan'' in
the title, ``school closure'' anywhere in the section field, a qualifying date, and the absence of
``workplace safety'' from the title. The returned result cards expose the actual section headings
of each matching webpage. The agent inspects those headings and fetches a named section only after
seeing which sections the webpage contains; it does not guess a section name during search.

Fielded Boolean queries remain central to professional search, including medical systematic
reviews \citep{wang2023mesh,wang2023chatgpt,wang2025reassessing,wang2026autobool}. A longstanding
barrier, however, is the effort and specialist knowledge required to formulate and revise them.
Language-model agents change this trade-off. They can construct compositional queries from a
user's question, inspect the returned evidence, and revise the constraints during the research
loop. The constraints remain explicit, while the agent takes on much of the formulation work.

We instantiate this idea in \textbf{Sieve}. During search, the agent formulates a Boolean Query
Language (BQL) expression that selects eligible webpages over their fields. An interchangeable
ranker then orders the selected set. During inspection, compact result cards expose titles, section
headings, and query-focused snippets. During fetching, the agent requests a named section rather
than visiting the complete webpage. Structure is therefore preserved across all three actions
rather than discarded between retrieval and evidence access. Appendices~\ref{app:grammar}--
\ref{app:bql-examples} describe the language, agent instructions, issued queries, and a complete
logged interaction.

Existing QA collections do not directly support this comparison because they do not provide
controlled access to webpage structure. We therefore derive paired flat and structured variants
of BrowseComp-Plus, HotpotQA, and MuSiQue. Within each pair, the questions and emitted text are
identical; the flat variant hides section and metadata fields, whereas the structured variant makes
them explicitly addressable. This design tests access to structure without changing the evidence
available to the agent.

Across agent strategies and ranker choices, \textsc{Sieve} achieves the highest accuracy on all
three collections. Its default BM25+Dense configuration improves over the most accurate
Search--Visit configuration by $1.6/3.1/0.7$ accuracy points while using $20.7$--$50.6\%$ fewer
tokens. Matched Search--Fetch controls show that BQL candidate selection improves all nine
ranker--collection pairs. Further analyses show that query-focused snippets make section-level
inspection effective and that the gains persist across retrievers and three agent backbones.

\paragraph{Contributions.}
This work makes three contributions. \textbf{First,} we identify the mismatch between structured
webpages and whole-webpage Search--Visit workflows, and formulate search--inspect--fetch as a
strategy for preserving structure from retrieval through evidence access. \textbf{Second,} we introduce
\textsc{Sieve}, which combines fielded Boolean candidate selection, interchangeable ranking,
structure-rich result inspection, and selective section fetching. \textbf{Third,} we show that
\textsc{Sieve} improves accuracy while using $20.7$--$50.6\%$ fewer tokens than the strongest
conventional Search--Visit configuration on each collection. Matched analyses further identify the
roles of BQL selection and query-focused snippets and show that the gains persist across retrievers
and agent backbones.

\section{Related Work}
\label{sec:related_work}

\textsc{Sieve} builds on three areas: iterative retrieval for research agents, Boolean query
formulation, and retrieval over webpage structure.

\paragraph{Retrieval workflows for research agents.}
Deep-search agents commonly use a ReAct-style loop \citep{yao2023react}: keyword or dense search
returns a top-$k$ list, after which the agent visits a webpage
\citep{nakano2021webgpt,jin2025searchr1,li2025searcho1,song2025r1searcher,sun2025zerosearch,gao2025asearcher}.
QA systems improve planning, memory, and evidence aggregation
\citep{trivedi2023ircot,gutierrez2024hipporag,gutierrez2025hipporag2}, but generally retain this
Search--Visit workflow. DCI instead removes the retriever and lets an agent search raw files
directly \citep{dci2026,subramanian2026keyword}, reporting cost advantages under cached file access and
aggregate pricing assumptions. RISE uses BM25 to bound the corpus in which an agent performs such
direct interaction \citep{rise2026}. We instead ask whether preserving webpage structure from
search through evidence access improves efficiency without sacrificing effectiveness.

\paragraph{Boolean search and agent-generated queries.}
Boolean retrieval expresses precise inclusion and exclusion criteria over structured records.
Field restrictions distinguish a term in a title from the same term in the body; Boolean operators
combine constraints; and phrases, wildcards, and ranges refine the candidate set. These capabilities
remain important in specialist search, particularly biomedical systematic reviews
\citep{wang2023mesh,wang2023chatgpt,wang2025reassessing,wang2026autobool}. This line of work shows
both why fielded Boolean queries remain useful and why their formulation is difficult: controlled
terminologies can improve retrieval, while query effectiveness depends strongly on the generating
model, prompt, and validation procedure. Once a Boolean query has selected a candidate pool, a
separate ranker can prioritize the webpages within it \citep{wang2023screening}. This separation
between selection and ranking is also central to \textsc{Sieve}. Earlier retrieval models combined exact selection with
graded evidence and field-sensitive ranking
\citep{turtle1991inference,metzler2004combining,robertson2004bm25f}. Most directly, \citet{clarke2026boolean}
equip an LLM-based agent with a Boolean retrieval engine over MS MARCO segments and obtain strong
first-stage retrieval effectiveness using only substring-match density. Their work supports Boolean
retrieval as an agent tool, but studies segment ranking rather than structured webpages or
the downstream search--inspect--fetch interaction. SIRA compiles an answer sketch into a weighted
BM25 query \citep{sira2026}, while LogicalRAG generates Boolean queries within a multi-turn loop
\citep{logicalrag2026}. Both still return whole webpages. \textsc{Sieve} instead connects fielded
candidate selection to result inspection and section fetching.

\paragraph{Structure and retrieval granularity.}
Focused retrieval has long targeted passages or webpage components
\citep{callan1994passage,trotman2005nexi}. More recent systems vary dense indexing granularity
\citep{chen2024densex} or construct hierarchies over otherwise unstructured text
\citep{sarthi2024raptor}. \textsc{Sieve} instead uses structure already available in the collection:
fielded constraints construct a candidate set, a separate ranker orders it, result listings expose
matching sections, and the fetch action returns a selected section. We ask whether carrying
webpage structure through this complete interaction improves both answer quality and context
efficiency. Appendix~\ref{app:engine} additionally compares the proposed executor with an
Indri-style structured-retrieval alternative while holding result cards and section fetching fixed.

\section{\textsc{Sieve}: Search--Inspect--Fetch}
\label{sec:method}

\textsc{Sieve} follows one design principle: webpage structure should survive the complete
interaction, from retrieval to evidence access. If search can identify a relevant section but the result
list discards that information, or if the subsequent visit still returns the whole webpage, the
structure provides little practical benefit. We therefore separate four decisions that
conventional Search--Visit workflows often collapse: which webpages qualify, how they are
ranked, what the agent sees before content access, and which content enters its context
(Figure~\ref{fig:architecture}).
Operationally, one turn moves from a BQL query to ranked result cards and then to a selected
section; the evidence returned from that section informs the next turn.

Throughout, \emph{visit} returns a complete webpage,
\emph{inspect} means viewing its result card, and \emph{fetch} returns one named section.

\paragraph{Structured source representation.}
We represent a webpage $d$ by its metadata $m_d$ and an ordered sequence of heading--content pairs:
\begin{equation}
 d=\left(m_d,\left\langle(h_{d,j},x_{d,j})\right\rangle_{j=1}^{n_d}\right).
\end{equation}
A conventional visit returns the concatenated webpage
$x_d=x_{d,1}\oplus\cdots\oplus x_{d,n_d}$, whereas a section fetch returns one selected
$x_{d,j}$. The representation preserves the emitted webpage text while separating titles,
headings, section bodies, and available metadata. Which fields are addressable depends on the
collection and operation: search uses indexed fields, while result cards and fetch may expose
additional metadata such as infoboxes. Appendix~\ref{app:corpus} gives the exact representation of
each collection.

The difference is not whether a heading remains visible in the text, but whether the tools let the
agent act on it. In a flat record, the agent cannot restrict a query to that heading or name its
section as a fetch target. This distinction motivates both the paired collections in
\S\ref{sec:controlled-collections} and the complete-system comparison.

\paragraph{Search: Boolean selection followed by ranking.}
\label{sec:method-search}
The agent submits a BQL expression that can combine terms with
\texttt{AND}, \texttt{OR}, and \texttt{NOT}; restrict terms to fields; use phrases and wildcards;
and filter date ranges. Its purpose is to control eligibility, not to replace ranking. For a
collection $\mathcal{D}$ and BQL expression $q$, the search result is
\begin{equation}
 \begin{aligned}
 \mathcal{C}_q &= \{d\in\mathcal{D}:d\models q\},\\
 \mathcal{L}_q &= \operatorname{TopK}_{d\in\mathcal{C}_q} R(d,q^+),
 \end{aligned}
 \label{eq:sieve-retrieval}
\end{equation}
where $q^+$ denotes the positive query terms and $R$ is an interchangeable ranker. This separation
is important: Boolean constraints express which sources are acceptable, while ranking decides
which acceptable sources the agent should inspect first.

The BQL expression compiles to a Lucene filter that constructs $\mathcal{C}_q$.
\textsc{Sieve}'s default $R$ combines BM25 scores over positive title and body terms with dense
scores from \textbf{BAAI/bge-base-en-v1.5} using reciprocal rank fusion
\citep{robertsonzaragoza2009bm25,xiao2023bge,cormack2009rrf}. Lexical and semantic scores provide
complementary orderings within the constrained set \citep{wang2021bert}. The ranking stage remains modular: our
controlled variants use BM25 or dense ranking alone, and no ranker can reintroduce a webpage
rejected by the Boolean filter.

Exact constraints are useful only if an over-constrained query can recover. If the filter admits no
webpages, \textsc{Sieve} relaxes the Boolean constraints. For a multi-constraint query, it returns
partial matches, prioritizing constraint coverage and applying the configured ranker $R$ within
each coverage tier. If no partial match exists, it ranks the full collection with BM25 over the
query's positive terms. The fallback is invoked only after a zero-result query and
never broadens a non-empty candidate set. It occurs in $36.8$--$53.7\%$ of \textsc{Sieve}'s search
calls. Its BM25 component is implemented separately from the Lucene scorer used for ordinary ranked
retrieval; Appendix~\ref{app:engine} quantifies their divergence. Together, precise selection and recovery let
the agent test a constrained hypothesis, obtain evidence when it is too restrictive, and
reformulate the next query.
Appendices~\ref{app:grammar}, \ref{app:bql-examples}, and~\ref{app:engine} give the grammar, issued
queries, compilation, and fallback implementation and validation.

\paragraph{Inspect: structure-rich result cards.}
A ranked list alone does not tell the agent why a webpage matched or which section is worth fetching;
returning the webpage body would remove the intended efficiency benefit. The top $k$ webpages are
therefore rendered as compact result cards. Each card identifies the source by title, exposes
section headings as possible fetch targets, and includes a $25$-token query-focused snippet as
local relevance evidence. To produce this snippet, the renderer removes Boolean operators and field
syntax, scores candidate passages by lexical overlap with the remaining positive terms, selects the
highest-scoring passage, and truncates it to $25$ whitespace-delimited tokens. Cards also expose
matched fields and available infobox keys. The snippet ablation changes only this representation:
candidate selection, ranking, and section fetching remain fixed.

\paragraph{Fetch: selective section access.}
The headings exposed during inspection become actionable through the fetch operation. The agent
chooses a listed webpage and requests one of its named sections or fields; the tool returns the
corresponding $x_{d,j}$ rather than the concatenated $x_d$. The agent may alternate search,
inspection, and fetching until it can answer or reaches the common interaction budget described in
\S\ref{sec:evaluation-protocol}. Fetch and visit share the same per-action token ceiling, so this
comparison changes access granularity rather than the maximum content available from one action.
Appendix~\ref{app:bql-examples-interaction} traces one complete logged interaction.

The complete \textsc{Sieve} strategy therefore combines Boolean candidate selection, modular
ranking, structure-rich result cards, and section-level fetching. Each component resolves a
different decision in the interaction: Boolean search controls admissibility, ranking prioritizes
candidates, cards support section selection, and fetch limits the resulting context. Our
default uses BM25+Dense ranking. We evaluate the complete strategy as our primary method, with
matched ablations that vary one component while holding the rest of the interaction fixed.

\section{Experimental Setup}
\label{sec:experimental_setup}

\subsection{Controlled Collections}
\label{sec:controlled-collections}
Existing QA benchmarks do not isolate access to webpage structure. We therefore derive paired flat
and structured versions of HotpotQA \citep{yang2018hotpotqa}, MuSiQue
\citep{trivedi2022musique}, and \textsc{BrowseComp-Plus} \citep{browsecompplus}. Within each pair,
the questions, webpages, and emitted text are identical. The flat version exposes title and body,
whereas the structured version additionally makes sections and available metadata explicitly
addressable.

The collections increase in difficulty and source complexity. HotpotQA is the more tractable
Wikipedia multi-hop setting; MuSiQue requires more connected composition; and
\textsc{BCP-S} poses difficult deep-search questions over a much larger, heterogeneous
web collection. We discuss results in this order. We report exact match on the Wikipedia
collections and LLM-judge accuracy on \textsc{BCP-S}, so accuracy should be compared
within rather than across collections.

For HotpotQA and MuSiQue, no model-based segmentation is needed. The Structured Wikipedia release
\citep{wikimedia2026structured} already provides sections and infoboxes. We match benchmark titles
to this collection and remove unmatched articles and questions left without gold evidence.
This retains $7{,}343/7{,}405$ HotpotQA and $2{,}409/2{,}417$ MuSiQue questions.
Appendix~\ref{app:corpus} gives the full preprocessing procedure.

\textsc{BrowseComp-Plus} contains $100{,}195$ webpages and $830$ questions but does not provide
addressable section annotations. We therefore use OpenAI's \texttt{gpt-5.5-nano} once to propose
section headings and boundaries. A deterministic procedure applies these boundaries without
rewriting the webpage text. Both variants contain the full collection and emit identical text,
and every experimental condition searches the same webpages.
Appendix~\ref{app:corpus} reports the full construction procedure and section-quality audit.

\paragraph{Why pair the collections?}
The paired design changes whether agents can act on webpage structure while holding the content
fixed. It does not remove every structural cue: headings remain visible in flat text, and section
boundaries can be recovered from many flat webpages. The contrast therefore measures the value of
exposing structure as addressable fields. The complete-system comparison asks the broader question
of whether \textsc{Sieve} improves deep search over conventional Search--Visit.

\subsection{Baseline Systems}
\label{sec:baseline-systems}

The baselines are organized around two decisions: whether retrieval is repeated and how much
content enters the agent's context after each search. Together with the matched no-BQL control,
they separate the value of iteration, selective content access, and BQL candidate selection instead of
attributing every difference to the complete system.

The one-shot \textbf{retrieve-then-read} baselines perform one retrieval and pass its results to one
answering call. \textbf{Search--AutoRead} retains an agent loop but returns full webpage text with
each search result, without a separate visit action. We run both with BM25 and dense retrieval.
The conventional \textbf{Search--Visit} baseline instead lets the agent search repeatedly, inspect
compact results, and visit selected webpages in full.

\textbf{Search--Fetch} retains iterative, unfiltered search but replaces whole-webpage visits with
section fetches. It is the closest no-Boolean control for \textsc{Sieve}: both use the same rankers,
top-$k$ result depth, and section fetching, while \textsc{Sieve} adds BQL candidate selection. We run Search--Visit, Search--Fetch, and
\textsc{Sieve} with BM25, dense, and BM25+Dense ranking.
These matched controls differ only in BQL selection, not in ranker, result depth, or access
granularity.

\textbf{DCI} gives the agent shell search and file-access tools over the shared experimental corpus
\citep{dci2026}. \textbf{BM25-bounded DCI} first stages a fixed working set of ten retrieved
webpages and exposes the same tools within the set, following RISE \citep{rise2026}.
Appendix~\ref{app:dci} describes these reimplementations and their differences from the published setups;
Appendix~\ref{app:skills} maps each condition to its instructions and reproduces the Search--Fetch manual.

\subsection{Models and Retrieval}
\label{sec:models-retrieval}

\paragraph{Agent backbones.}
The primary agent is \textbf{Tongyi-DeepResearch-30B-A3B} \citep{tongyi2025}, served with vLLM
\citep{kwon2023vllm} in its released ReAct scaffold. We preserve that scaffold and vary only the
collection-access tools and their condition-specific instructions. The agent-backbone study
additionally uses \textbf{Qwen-AgentWorld-35B-A3B} \citep{zuo2026qwenagentworld} and
\textbf{OpenResearcher-30B-A3B} \citep{li2026openresearcher}. Every generation uses temperature
$0.6$ and seed $42$. Serving details are in Appendix~\ref{app:setup}.

\paragraph{Retrievers.}
Sparse retrieval uses Pyserini/Lucene BM25 \citep{pyserini2021,anserini}; dense retrieval uses exact
inner-product search with \textbf{BAAI/bge-base-en-v1.5}, and BM25+Dense combines the two rankings
by reciprocal rank fusion \citep{cormack2009rrf}. In \textsc{Sieve}, each ranker operates only on
the BQL-admitted candidate set. To test retriever sensitivity, we replace only \textsc{Sieve}'s
dense channel with the small, base, and large bge-en-v1.5 models \citep{xiao2023bge}, and
Qwen3-Embedding-0.6B, 4B, and 8B \citep{zhang2025qwen3embedding}, while holding all other
components and hyperparameters fixed.

\subsection{Evaluation Protocol}
\label{sec:evaluation-protocol}

\paragraph{Budgets.}
Every search call returns at most $k{=}5$ results, every visit or fetch has a $12{,}000$-token ceiling, and
every iterative condition has a maximum of $100$ agent steps. The ten-webpage DCI working set is
a one-time staging depth and is separate from the per-call $k$ used by search agents.

\paragraph{Metrics.}
We use the standard accuracy metric for each collection. On
\textsc{BCP-S}, accuracy is the verdict from OpenAI's \texttt{gpt-4o-mini}
judge; on HotpotQA and MuSiQue, it is SQuAD-style exact match.
We refer to these benchmark-specific measures collectively as
\emph{accuracy}. \emph{Tok.} is the mean total number of input and output tokens used to answer a
question, without recounting earlier conversation history at later calls; \emph{LLM calls} is the
mean number of model invocations. Appendix~\ref{app:fulltables}
also reports step-summed tokens, which recount the growing conversation history at every call.
The primary token measure captures how much distinct content the strategy introduces, while the
step-summed measure better reflects serving work when the full conversation is processed again.
Reporting both prevents a reduction caused only by one accounting convention.
All accuracy comparisons are paired on identical questions. We use exact McNemar tests for binary outcomes and paired $t$-tests for tokens and calls, two-sided at $\alpha{=}0.05$, with Bonferroni correction.

\paragraph{Implementation.}
Our implementation of \textsc{Sieve}, including the BQL executor, search and fetch tools,
collection builders, and evaluation harness, is included in the SkimSearchAgent library at
\url{https://github.com/ielab/skim-search-agent}.

\section{Results}
\label{sec:results}

Table~\ref{tab:consolidated} presents the full system comparison. We first ask whether
\textsc{Sieve} improves the accuracy--context trade-off over conventional Search--Visit, then use
the remaining systems to identify where that improvement comes from. Complete per-collection
metrics are reported in Appendix~\ref{app:fulltables}.

\begin{table*}[t]
\centering
\footnotesize
\setlength{\tabcolsep}{1.7pt}
\renewcommand{\arraystretch}{0.95}
\caption{Main system comparison. \emph{Acc.} is exact match on HotpotQA and MuSiQue and
the LLM-judge on \textsc{BCP-S}; \emph{Tok.} and \emph{LLM calls} report mean context use
and model invocations per question. Bold marks the best iterative-system value in each column.
Superscripts $a$ and $b$ denote significance versus BM25 Search--Visit and the default
\textsc{Sieve}, respectively, after Bonferroni correction. Full metrics are in
Appendix~\ref{app:fulltables}.}
\label{tab:consolidated}
\begin{tabular*}{\textwidth}{@{\extracolsep{\fill}}>{\raggedright\arraybackslash}p{4.8cm}rrr|rrr|rrr@{}}
\toprule
& \multicolumn{3}{c}{HotpotQA}
& \multicolumn{3}{c}{MuSiQue}
& \multicolumn{3}{c}{\textsc{BCP-S}} \\
\cmidrule(lr){2-4}\cmidrule(lr){5-7}\cmidrule(lr){8-10}
Condition & Acc. & Tok. & LLM calls
& Acc. & Tok. & LLM calls
& Acc. & Tok. & LLM calls \\
\midrule
\multicolumn{10}{l}{\textit{One-shot retrieve-then-read}} \\
BM25 & 29.9$^{ab}$ & 1k$^{ab}$ & 1.0$^{ab}$ & 7.8$^{ab}$ & 2k$^{ab}$ & 1.0$^{ab}$ & 2.2$^{ab}$ & 2k$^{ab}$ & 1.0$^{ab}$ \\
Dense & 30.3$^{ab}$ & 1k$^{ab}$ & 1.0$^{ab}$ & 11.0$^{ab}$ & 1k$^{ab}$ & 1.0$^{ab}$ & 6.9$^{ab}$ & 2k$^{ab}$ & 1.0$^{ab}$ \\
\midrule
\multicolumn{10}{l}{\textit{Iterative Search--AutoRead}} \\
BM25 & 42.4$^{b}$ & 72k$^{ab}$ & 36.8$^{ab}$ & 20.9$^{ab}$ & 108k$^{ab}$ & 78.3$^{ab}$ & 13.6$^{ab}$ & 142k$^{ab}$ & 85.4$^{ab}$ \\
Dense & 39.9$^{ab}$ & 62k$^{ab}$ & 35.8$^{ab}$ & 20.5$^{ab}$ & 95k$^{ab}$ & 71.2$^{ab}$ & 13.0$^{ab}$ & 122k$^{ab}$ & 76.3$^{ab}$ \\
\midrule
\multicolumn{10}{l}{\textit{Direct corpus interaction}} \\
DCI (no retriever) & 42.8$^{b}$ & 35k$^{ab}$ & 21.8$^{a}$ & 26.9 & 63k$^{ab}$ & 39.4$^{ab}$ & 22.2$^{ab}$ & 107k$^{ab}$ & \textbf{53.6}$^{ab}$ \\
BM25-bounded DCI (RISE-style) & 42.6$^{b}$ & 21k$^{ab}$ & 22.1$^{ab}$ & 26.3$^{b}$ & 47k$^{ab}$ & 39.6$^{ab}$ & 33.4 & 75k$^{ab}$ & 64.0 \\
\midrule
\multicolumn{10}{l}{\textit{Iterative Search--Visit (whole-webpage visits)}} \\
BM25 & 43.7$^{b}$ & 20k$^{b}$ & \textbf{15.9}$^{b}$ & 26.1$^{b}$ & 43k$^{b}$ & \textbf{28.7}$^{b}$ & 34.7 & 68k$^{b}$ & 64.1 \\
Dense & 43.0$^{b}$ & 21k$^{ab}$ & 25.9$^{ab}$ & 25.5$^{b}$ & 42k$^{b}$ & 47.6$^{ab}$ & 36.5 & 58k$^{ab}$ & 65.7 \\
BM25+Dense & 40.2$^{ab}$ & 21k$^{ab}$ & 35.4$^{ab}$ & 21.8$^{ab}$ & 39k$^{ab}$ & 56.3$^{ab}$ & 33.7 & 53k$^{ab}$ & 68.5$^{b}$ \\
\midrule
\multicolumn{10}{l}{\textit{Iterative Search--Fetch (section fetching)}} \\
BM25 & 41.0$^{ab}$ & \textbf{13k}$^{ab}$ & 31.1$^{ab}$ & 24.5$^{b}$ & 22k$^{ab}$ & 50.3$^{ab}$ & 30.6$^{b}$ & 40k$^{ab}$ & 76.6$^{ab}$ \\
Dense & 40.7$^{ab}$ & 14k$^{a}$ & 33.2$^{ab}$ & 23.4$^{ab}$ & 23k$^{ab}$ & 53.9$^{ab}$ & 30.1$^{b}$ & 37k$^{ab}$ & 78.6$^{ab}$ \\
BM25+Dense & 40.5$^{ab}$ & 14k$^{a}$ & 32.6$^{ab}$ & 24.6$^{b}$ & 23k$^{ab}$ & 53.0$^{ab}$ & 34.9 & \textbf{35k}$^{ab}$ & 73.6$^{ab}$ \\
\cmidrule(lr){1-10}
\textsc{Sieve} (BQL-filtered BM25) & 44.5 & 15k$^{ab}$ & 27.9$^{ab}$ & 27.9 & 24k$^{ab}$ & 47.6$^{ab}$ & 33.6 & 49k$^{a}$ & 62.0 \\
\textsc{Sieve} (BQL-filtered Dense) & \textbf{45.8}$^{a}$ & 15k$^{ab}$ & 27.9$^{ab}$ & 28.2 & 24k$^{ab}$ & 48.0$^{ab}$ & 34.6 & 48k$^{a}$ & 61.5 \\
\textsc{Sieve} (BQL-filtered BM25+Dense) & 45.3$^{a}$ & 14k$^{a}$ & 20.9$^{a}$ & \textbf{29.2}$^{a}$ & \textbf{21k}$^{a}$ & 32.6$^{a}$ & \textbf{37.2} & 46k$^{a}$ & 62.2 \\
\bottomrule
\end{tabular*}
\end{table*}

\subsection{Accuracy and Context Efficiency}
\label{sec:results-efficiency}

\textsc{Sieve} improves accuracy without paying for that improvement with more context. Against
the most accurate Search--Visit configuration on each collection, its default BM25+Dense setting
improves accuracy by $1.6$, $3.1$, and $0.7$ points on HotpotQA, MuSiQue, and \textsc{BCP-S}, while
using $30.4\%$, $50.6\%$, and $20.7\%$ fewer tokens, respectively. The gains over BM25
Search--Visit are significant on HotpotQA and MuSiQue. On \textsc{BCP-S}, judge accuracy and the
stricter exact-match measure agree: both improve, by $2.5$ and $3.4$ points over BM25
Search--Visit, respectively.

\finding{1}{Across all three collections, \textsc{Sieve} improves accuracy over the strongest
conventional Search--Visit configuration while using $20.7$--$50.6\%$ fewer tokens.}

The reduction comes from controlling what enters the context, rather than curtailing the search.
\textsc{Sieve} makes more model calls on the Wikipedia collections and a similar number on
\textsc{BCP-S}, but receives substantially less text overall. Section-level
fetching lets the agent continue following an evidence chain without repeatedly carrying complete
webpages into later decisions. Step-summed token accounting leads to the same conclusion
(Appendix~\ref{app:fulltables}).
The number of calls alone is therefore a poor proxy for context cost. \textsc{Sieve} can take an
additional search or fetch step while keeping each step focused on a much smaller portion of the
collection.

\begin{figure}[H]
\centering
\includegraphics[width=0.56\linewidth]{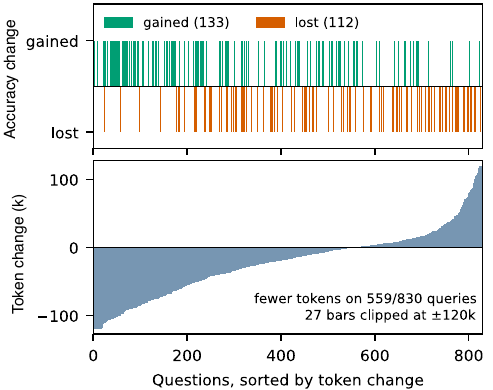}
\caption{Per-question changes from BM25 Search--Visit to \textsc{Sieve} on \textsc{BCP-S}.
Both panels share the ordering induced by token change. \emph{Top}: questions answered correctly
only by \textsc{Sieve} (gained) or only by Search--Visit (lost); unchanged outcomes lie on the
centre line. \emph{Bottom}: change in distinct tokens
(\textsc{Sieve} minus Search--Visit), clipped at $\pm120$k for display.~\vspace{-5pt}}
\label{fig:gainloss}
\end{figure}

Figure~\ref{fig:gainloss} shows that the aggregate saving reflects a collection-wide shift rather
than a few unusually long webpages: \textsc{Sieve} uses fewer tokens on roughly two-thirds of
\textsc{BCP-S} questions. More importantly, accuracy gains cluster on the token-saving side of the
plot, while losses become more common as \textsc{Sieve} consumes more. Efficiency and effectiveness
are therefore aligned rather than traded off. Selective access appears to improve not only how
much context the agent receives, but also its quality, by keeping distracting webpage content out
of evidence selection and synthesis.
The broader accuracy--context frontier supports the same conclusion: configurations that consume more context
do not systematically answer more questions correctly (Appendix Figure~\ref{fig:pareto}).

\subsection{Full-System Comparison}
\label{sec:results-comparison}

\paragraph{Iteration must be selective.}
The one-shot baselines show that difficult questions require the agent to revise its search as
evidence accumulates. Search--AutoRead shows that iteration alone is insufficient: adding every
result in full increases context without a comparable accuracy gain. Effective deep search therefore
requires repeated search and selective access.

\paragraph{Corpus access still needs retrieval.}
DCI is competitive on the Wikipedia collections but weaker on the larger, heterogeneous
\textsc{BCP-S}; BM25 bounding recovers much of that gap. Direct corpus tools therefore do not
replace retrieval: candidates must still be organized into a manageable evidence space.
Appendix~\ref{app:dci} describes our implementation and its differences from the published setup.

\paragraph{Ranking and access are coupled.}
The best ranker changes across collections and access strategies. Search--Visit ranks the whole
collection, whereas \textsc{Sieve} ranks within a BQL-eligible set. Ranking quality must therefore
be assessed together with downstream access, not in isolation.

\paragraph{BQL adds candidate selection.}
Search--Fetch is the matched no-BQL control: it uses the same rankers and section fetching but
without BQL eligibility constraints (\S\ref{sec:baseline-systems}). \textsc{Sieve} improves accuracy
in all nine pairs, showing that the gain is not tied to one ranker. BQL instead adds a complementary
decision: it determines what may be considered, while ranking determines what should be inspected
first. Section~\ref{sec:ablation-bql-use} examines how agents use this separation.

\section{Ablation and Diagnostic Analysis}
\label{sec:ablation}

Having compared the complete systems, we now ask which components produce the gain,
whether it persists across retrievers and agent backbones, and where errors remain. Additional
controls appear in Appendix~\ref{app:controls}.

\subsection{Components and Robustness}
\label{sec:ablation-components}

\begin{figure}[H]
\centering
\includegraphics[width=0.6\linewidth]{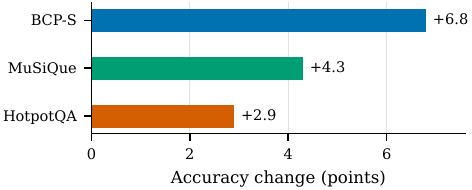}
\caption{Accuracy change from adding query-focused snippets. Full accuracy and cost values appear
in Appendix Table~\ref{tab:snippet-ablation}.}
\label{fig:snippet-ablation}
\end{figure}

\paragraph{Query-focused snippets.}
An addressable section is useful only if the agent can judge whether it is worth fetching. A
heading describes the section's topic, but may not reveal its relevance to the current question.
We therefore remove the $25$-token query-focused snippet while keeping candidate selection,
ranking, instructions, result depth, and fetching unchanged
(Figure~\ref{fig:snippet-ablation}).

Removing snippets lowers accuracy by $2.9$--$6.8$ points across the three collections, with a
significant drop in every case. The agent also makes more model calls, but this additional search
does not recover the lost accuracy. The snippets make the exposed structure actionable: they give
the agent enough local evidence to decide which section should be fetched next.

\finding{2}{Query-focused snippets raise accuracy by $2.9$--$6.8$ points across all three
collections, showing that exposed structure must also be actionable.}

\begin{figure}[H]
\centering
\includegraphics[width=0.8\linewidth]{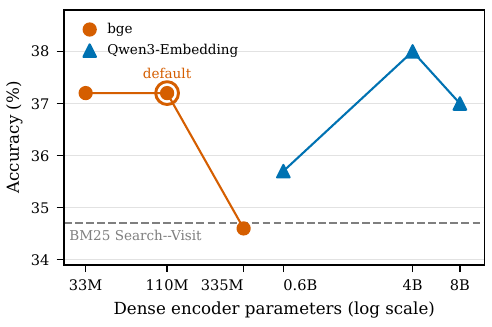}
\caption{Accuracy by dense-encoder size on \textsc{BCP-S}. Lines distinguish encoder
families, the outer circle marks the default encoder, and the dashed line marks BM25
Search--Visit. Full accuracy and cost measurements appear in Appendix
Table~\ref{tab:retriever-sensitivity}.}
\label{fig:retriever-sensitivity}

\end{figure}

\paragraph{Retriever choice.}
BQL determines the eligible webpages, while the ranker determines which candidates the agent sees
first. To separate these roles, we replace only the dense retriever in \textsc{Sieve} and keep the
rest of the system fixed. We run this analysis on
\textsc{BCP-S}, where retrieval accounts for most remaining errors
(Figure~\ref{fig:retriever-sensitivity}).

Context use remains between $46$k and $48$k tokens for all six encoders, compared with $68.1$k for
Search--Visit, while accuracy ranges from $34.6$ to $38.0$. Qwen3-Embedding-4B
\citep{zhang2025qwen3embedding} performs best. Larger encoders tend to achieve higher accuracy,
but the trend is not monotonic: Qwen3-Embedding-8B trails 4B, and bge-large trails the smaller bge
models. Scale can therefore help, but does not determine retrieval quality. The narrow context
range shows that the saving is not tied to a particular encoder.

\begin{figure*}[t]
\centering
\includegraphics[width=\textwidth]{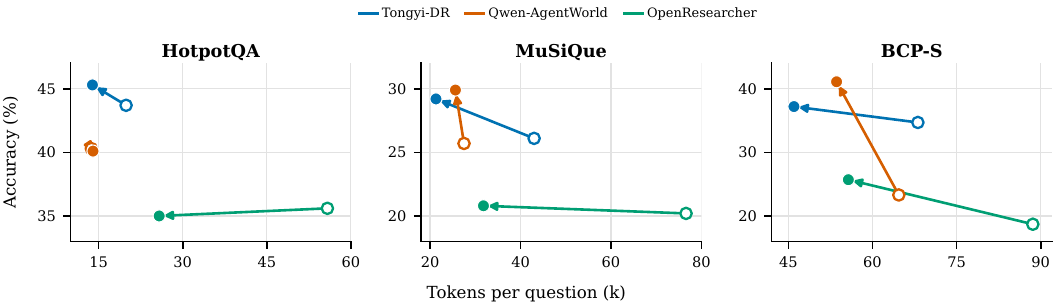}
\caption{Agent-backbone transfer across all three collections. Arrows run from BM25 Search--Visit
(open) to \textsc{Sieve} (filled); axes are scaled separately for readability. Appendix
Table~\ref{tab:backbone-transfer} reports the exact values.}
\label{fig:agent-backbone}
\end{figure*}

\paragraph{Agent backbone.}
The agent backbone controls query reformulation, result inspection, and section fetching. We
repeat the Search--Visit and \textsc{Sieve} comparison with two additional deep-search agents to
test whether the result depends on Tongyi-DeepResearch. Across the nine backbone--collection
pairs, \textsc{Sieve} uses less context in eight and improves accuracy in seven; the remaining two
accuracy differences are within $0.6$ points (Figure~\ref{fig:agent-backbone}). On
\textsc{BCP-S}, all three backbones move toward both higher accuracy and lower
context use.

The largest savings occur when the baseline consumes substantial context. With OpenResearcher, \textsc{Sieve}
more than halves token use on both Wikipedia collections while keeping accuracy within $0.6$
points. Both alternative agents also make substantial accuracy gains on
\textsc{BCP-S} with less context. The complete results appear in Appendix
Table~\ref{tab:backbone-transfer}. These transfers show that the main result is not tied to one
agent's search policy or prompting conventions.

\finding{3}{Across three agent backbones, \textsc{Sieve} uses less context in eight of nine
backbone--collection pairs and improves accuracy in seven.}

\subsection{Failures and BQL Use}
\label{sec:ablation-failures}

\begin{figure*}[t]
\centering
\begin{tikzpicture}
\begin{axis}[
    xbar stacked,
    width=0.90\textwidth,
    height=4.8cm,
    xmin=0,
    xmax=100,
    bar width=6.5pt,
    xlabel={Share of all questions (\%)},
    xlabel style={font=\small},
    xtick={0,25,50,75,100},
    xticklabel style={font=\scriptsize},
    symbolic y coords={
        {\textsc{BCP-S} / \textsc{Sieve}},
        {\textsc{BCP-S} / Search--Visit},
        {MuSiQue / \textsc{Sieve}},
        {MuSiQue / Search--Visit},
        {HotpotQA / \textsc{Sieve}},
        {HotpotQA / Search--Visit}
    },
    ytick=data,
    yticklabel style={font=\scriptsize},
    axis y line*=left,
    axis x line*=bottom,
    tick align=outside,
    major grid style={draw=black!12},
    xmajorgrids=true,
    legend style={
        at={(0.5,1.03)},
        anchor=south,
        legend columns=4,
        draw=none,
        font=\tiny,
        /tikz/every even column/.append style={column sep=3pt}
    },
]
\addplot+[draw=white, fill=cbBlue]
coordinates {
    (4.3,{MuSiQue / \textsc{Sieve}})
    (4.6,{MuSiQue / Search--Visit})
    (4.1,{HotpotQA / \textsc{Sieve}})
    (4.1,{HotpotQA / Search--Visit})
    (42.7,{\textsc{BCP-S} / \textsc{Sieve}})
    (46.5,{\textsc{BCP-S} / Search--Visit})
};
\addlegendentry{Retrieval}
\addplot+[draw=white, fill=cbOrange]
coordinates {
    (6.4,{MuSiQue / \textsc{Sieve}})
    (9.2,{MuSiQue / Search--Visit})
    (2.2,{HotpotQA / \textsc{Sieve}})
    (4.0,{HotpotQA / Search--Visit})
    (5.4,{\textsc{BCP-S} / \textsc{Sieve}})
    (7.9,{\textsc{BCP-S} / Search--Visit})
};
\addlegendentry{Selection}
\addplot+[draw=white, fill=cbGray]
coordinates {
    (60.1,{MuSiQue / \textsc{Sieve}})
    (60.1,{MuSiQue / Search--Visit})
    (48.4,{HotpotQA / \textsc{Sieve}})
    (48.2,{HotpotQA / Search--Visit})
    (17.4,{\textsc{BCP-S} / \textsc{Sieve}})
    (14.5,{\textsc{BCP-S} / Search--Visit})
};
\addlegendentry{Synthesis}
\addplot+[draw=white, fill=cbBluishGreen]
coordinates {
    (29.2,{MuSiQue / \textsc{Sieve}})
    (26.1,{MuSiQue / Search--Visit})
    (45.3,{HotpotQA / \textsc{Sieve}})
    (43.7,{HotpotQA / Search--Visit})
    (34.5,{\textsc{BCP-S} / \textsc{Sieve}})
    (31.1,{\textsc{BCP-S} / Search--Visit})
};
\addlegendentry{Correct}
\end{axis}
\end{tikzpicture}
\caption{Outcome composition over all questions. Retrieval means no gold webpage surfaced;
selection means one surfaced but no content was accessed; synthesis means an incorrect answer
after accessing one. Correctness uses exact match, including on \textsc{BCP-S}.}
\label{fig:failure-decomposition}
\end{figure*}
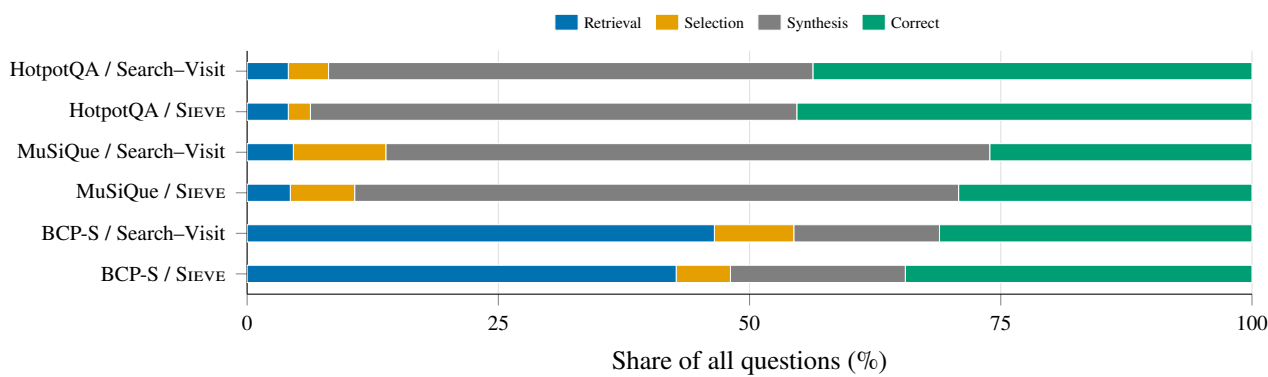

\paragraph{Where errors arise.}
Using the interaction traces, we assign each question one of four outcomes: correct, retrieval
failure, selection failure, or synthesis failure (Figure~\ref{fig:failure-decomposition}).

\textsc{Sieve} answers more questions correctly and reduces selection failures on every
collection, consistent with structured cards helping the agent choose which evidence to fetch.
The main bottleneck differs by collection: retrieval failures account for
$42.7$--$46.5\%$ of \textsc{BCP-S} questions, whereas synthesis failures account for
$48.2$--$60.1\%$ on the Wikipedia collections. Synthesis failures remain comparatively stable
between Search--Visit and \textsc{Sieve}, suggesting that errors after evidence access depend more
on the agent's extraction and reasoning capabilities than on the retrieval strategy.
Further gains therefore require better retrieval on \textsc{BCP-S}, but better evidence
extraction and synthesis on HotpotQA and MuSiQue. Appendix~\ref{app:traces} provides examples of
all three failure types.

\paragraph{How agents use BQL.}
\label{sec:ablation-bql-use}
The logs confirm that agents use BQL's added expressivity. Field restrictions
appear in $42.8$--$69.1\%$ of questions, and Boolean combinations in $34.1$--$44.1\%$
(Appendix Table~\ref{tab:operator-adoption}); representative queries are given in
Appendix~\ref{app:bql-examples}. Exact constraints and fallback retrieval play complementary roles.
Calls with matches retain the submitted constraints, while a zero-hit call triggers ranked term
retrieval. This fallback handles $36.8$--$53.7\%$ of search calls, and removing it lowers
\textsc{BCP-S} accuracy by $6.4$ points. The fallback is therefore a necessary part of the
Boolean-capable design, not a replacement for Boolean selection: non-empty queries preserve the
agent's explicit constraints, while recovery lets it revise an over-constrained intermediate
query using newly surfaced terms. Together with the nine gains over the ranker-matched no-BQL
control (\S\ref{sec:results-comparison}), this shows that precise selection and reliable recovery
are complementary. Additional controls show that the gain cannot be explained by structure
exposure alone or reproduced by a matched Indri-style executor
(Appendices~\ref{app:controls} and~\ref{app:engine}).

\section{Conclusion}
\label{sec:conclusion}
In this study, we exploit webpage structure to improve how deep-search agents find and access
evidence. We introduce \textsc{Sieve}, a search--inspect--fetch strategy built around fielded
Boolean retrieval. A \textsc{Sieve} configuration achieves the highest observed accuracy on each
of three QA collections; its default also outperforms the best conventional Search--Visit
configuration while using $20.7$--$50.6\%$ fewer tokens. These gains persist across retrievers and
three agent backbones.

Our findings show that source structure is more than a formatting or preprocessing detail.
Exposing it as searchable and fetchable fields shapes what an agent retrieves, inspects, and
includes in its context. This points to a broader direction: deep-search retrieval should preserve source
organization rather than flatten webpages, allowing agents to access information at its original
granularity.

Language-model agents also change the practical role of Boolean retrieval. Historically, authoring
and revising precise fielded queries required specialist effort. An agent can perform that work
during its research loop while retaining the explicit selection constraints that make Boolean
search useful. This makes capabilities long used in professional search practical for general
deep-search systems.

\section*{Limitations}

The comprehensive comparison uses one primary backbone and three QA collections. Transfer
experiments with Qwen-AgentWorld and OpenResearcher cover all three collections, but do not
establish generalization to other agents, domains, or live web search. Because judge overlays are
incomplete for the Wikipedia conditions, HotpotQA and MuSiQue use exact match as their primary
metric.

The experimental conditions are not fully symmetric. Query-language systems receive
condition-specific manuals, final-answer recovery is incomplete for some Wikipedia cells, and a
context-window early-stop rule was introduced part-way through the experiments.
Appendices~\ref{app:skills}, \ref{app:fairness}, and~\ref{app:setup} reproduce the instructions
and document these differences rather than assuming equivalence.

The structured/flat pairs hide addressable fields but retain their text and visible headings, so
they diagnose field access rather than the presence of structure itself. \textsc{Sieve} also
jointly changes candidate selection, result cards, and content access. Our controls isolate several
components, but do not fully cross every interaction; the evidence therefore supports the complete
system rather than an independent effect of each Boolean operator. Zero-hit BQL queries further
use an approximate BM25 implementation that differs from the Lucene ranker. \textsc{Sieve} is
thus a Boolean-capable workflow with designed recovery, not a strict Boolean or pure Lucene system.

The \textsc{BCP-S} section trees are model-generated and may contain imperfect headings or
boundaries. Wikipedia articles are matched by title against a newer snapshot without a content
check, so article drift may add noise.

\paragraph{Ethical considerations.}
Our experiments use released QA benchmarks, collect no user data, and involve no human-subjects research.
Fielded constraints and selective fetching can nevertheless exclude relevant sources or remove
surrounding context, particularly when metadata or model-generated section boundaries are wrong.
Deployed systems should retain source provenance, permit full-webpage visits, and respect
access, privacy, and licensing restrictions. Lower token use should not be treated as evidence that
the retrieved material is complete or correct.

\bibliographystyle{plainnat}
\bibliography{custom}

\appendix
\section{Dataset Construction and Statistics}
\label{app:corpus}

\paragraph{\textsc{BrowseComp-Plus} webpage collection.}
We load and de-obfuscate the complete \path{Tevatron/browsecomp-plus} release: $100{,}195$
webpages and all $830$ questions. We preserve the released webpage identifiers, questions, and
relevance annotations. The flat and structured variants contain the same full webpage collection,
and every \textsc{BCP-S} condition searches it. Corpus membership is therefore
independent of the evaluation questions; no query-pool filtering or query-dependent webpage
selection is applied.

\paragraph{\textsc{BCP-S} fields and sectioning.}
Title, author, and date are parsed from the de-obfuscated webpage frontmatter; placeholder and
boilerplate author values are removed. A one-time pass with OpenAI's \texttt{gpt-5.5-nano}
proposes section headings and boundary lines. A deterministic procedure applies those boundaries;
the model does not rewrite webpage text. The sectioning requests cover all $100{,}195$ webpages.
Rare webpages that exceed the per-request input limit are truncated to fit before their boundaries
are generated. Title, cleaned author, date, and sections are scopeable in the structured
variant. The flat twin contains the same emitted text, including inserted headings and folded
author/date text, but hides those fields from fielded retrieval.

In a $20{,}000$-webpage audit sample, every webpage has a non-empty
\texttt{sections} field, $94.9\%$ have multiple sections, and the mean is $14.7$ sections per
webpage.

\paragraph{Wikipedia reconstruction.}
HotpotQA and MuSiQue are rebuilt from the Structured Wikipedia release
\citep{wikimedia2026structured}
(\texttt{enwiki\_namespace\_0}), which supplies native sections, infoboxes, and abstracts; no LLM
pass is used. We match benchmark titles to article names
and URL-derived titles after Unicode, case, punctuation, whitespace, and diacritic normalization.
Because the title is treated as the article identity, a title match is accepted without a content
check. Articles with no match, typically because they were renamed or deleted, are dropped. Gold references
to them are removed, followed by any question left with no gold webpage.

\begin{table}[ht]
\centering
\caption{Wikipedia collection sizes after title matching. \emph{Questions retained} reports
retained/original questions; \emph{Coverage} is the corresponding retention rate.~\vspace{-5pt}}
\label{tab:wikipedia-preprocessing}
\small
\setlength{\tabcolsep}{3pt}
\begin{tabular*}{\columnwidth}{@{\extracolsep{\fill}}lrrr@{}}
\toprule
Collection & Webpages & Questions retained & Coverage \\
\midrule
HotpotQA & $59{,}833$ & $7{,}343/7{,}405$ & $99.16\%$ \\
MuSiQue & $16{,}583$ & $2{,}409/2{,}417$ & $99.67\%$ \\
\bottomrule
\end{tabular*}
\end{table}

Each retained collection is emitted as two twins. The structured version retains title, sections,
and infobox data for result cards and fetching; the flat version exposes title and text only. Their webpage identifiers,
questions, qrels, and emitted text are identical, so BM25 and dense retrieval see the same content.
Fielded retrieval can scope over sections.
In a $20{,}000$-webpage HotpotQA sample, every webpage
has at least one section, $89.6\%$ have multiple sections, and the mean is $8.44$; a full scan of
$59{,}833$ webpages gives $89.3\%$ and $8.29$. Across all $16{,}583$ MuSiQue webpages, the
corresponding values are $100\%$, $87.9\%$, and $9.70$. These counts include the leading
\texttt{(intro)} unit. Unlike the Wikipedia structures, the \textsc{BCP-S} section trees
are model-generated, which we treat as a limitation.

\paragraph{Section-quality audit.}
Section counts alone do not show that the boundaries are useful. We therefore examine $200$ random
\textsc{BCP-S} webpages (seed 42). The sample averages $16.2$ sections per
webpage (median $9.0$; 10th--90th percentile $3.0$--$28.1$), and $98.0\%$ contain multiple
sections. The gap between the mean and median reflects a long right tail.
All sampled section lists partition their source text cleanly. Among headings for which the measure
is applicable, $87.1\%$ share a content word with their section body. Near-empty sections account
for $1.9\%$ of the sample. Repeated headings over different content occur in $3.5\%$ of webpages;
in the worst case, one heading is repeated across 83 sections. Thus, section fetching is
well supported for most webpages, although a small degenerate tail remains.

\section{Method Details}
\label{app:method-details}

This appendix specifies the BQL language and tool behavior used by \textsc{Sieve}. We first describe the query
grammar and executor behavior, then reproduce the instructions shown to the agents, and finally
give issued queries and a complete worked interaction.

\subsection{Query Grammar and Tool Behavior}
\label{app:grammar}

The agent-facing BQL supports field restrictions, Boolean composition, wildcards, phrases,
and date ranges. A query of the form \texttt{term[field]} restricts a term to a named field;
\texttt{AND}, \texttt{OR}, and \texttt{NOT} combine conditions; parentheses control grouping;
\texttt{word*} expresses a wildcard; and a quoted \texttt{"phrase"} requires an exact span. Where a
typed date field exists, \texttt{date[RANGE]} filters by date.

Available search fields depend on the corpus. The Wikipedia collections index title, section, and
body; infobox data are available for result-card inspection and fetching. \textsc{BCP-S} indexes title,
section, body, author, and date, but has no infobox field. Its webpages contain section trees;
$94.9\%$ of the $20{,}000$ sampled webpages are multi-section, and \texttt{fetch} can address
those sections.

When a hard \texttt{AND} query returns no exact matches, the default executor ranks webpages using
the query's positive terms. This fallback was enabled in every primary experiment, so the main
BQL-filtered conditions are not strict Boolean retrievers. An appendix control disables the
fallback while holding the remaining \textsc{Sieve} workflow fixed
(Table~\ref{tab:additional-controls}).

The lower-level executor also implements \texttt{NEAR/}$w$ with Lucene
\texttt{SpanNearQuery}, but the agent-facing tokenizer has no \texttt{NEAR} token. It rewrites
\texttt{NEAR/5(a,b)} into a string that the parser rejects. Proximity use is therefore zero by
construction. Two other parser behaviors were absent from the agent manuals. First, an unquoted
multi-word sequence compiles to an \texttt{AND} over its words rather than to the adjacent phrase
described in the manuals. Second, lowercase English \texttt{and}, \texttt{or}, and \texttt{not} are
parsed as Boolean operators even inside an otherwise plain query.
The language follows familiar Boolean and fielded-search conventions. For example, its field
restriction corresponds to \texttt{x[ti]} in Ovid/PubMed or \texttt{TI(x)} in Westlaw/Lexis;
Boolean operators map directly; and \texttt{word*} mirrors truncation forms such as Ovid's
\texttt{auth*}.

\subsection{Query-Language Instructions}
\label{app:skills}

\subsubsection{Instruction Assignment by Condition}
\label{app:skills-mapping}

Only conditions that expose a query language receive query-writing instructions. \textsc{Sieve}
ranker configurations receive the Search--Fetch instructions reproduced below, while the
appendix-only Indri comparison receives instructions for Indri syntax. BM25, dense, BM25+Dense,
DCI, Search--Visit, Search--Fetch, and retrieve-then-read receive no query-language manual.

\begin{table}[H]
\centering
\caption{Query-language manual assigned to each condition. ``Wikipedia'' denotes HotpotQA and
MuSiQue, and ``BrowseComp'' denotes \textsc{BCP-S}.}
\label{tab:skills-map}
\small
\setlength{\tabcolsep}{3pt}
\begin{tabular*}{\columnwidth}{@{\extracolsep{\fill}}>{\raggedright\arraybackslash}p{8.5cm}>{\raggedright\arraybackslash}p{12cm}@{}}
\toprule
Condition & Instructions shown to the agent \\
\midrule
All \textsc{Sieve} ranker configurations
& Search--Fetch manual for Wikipedia or BrowseComp \\
\midrule
Indri and Indri+Dense (appendix only)
& Indri query manual \\
\midrule
All other reported conditions
& \emph{None} \\
\bottomrule
\end{tabular*}
\end{table}

\subsubsection{Complete Search--Fetch Instructions}
\label{app:skills-full}

\textsc{Sieve} receives the instructions below on HotpotQA and MuSiQue. Its ablation without dense
fusion receives identical instructions. Line breaks are wrapped at 80 columns for typesetting;
the wording is otherwise unchanged. Section~\ref{app:skills-diffs} explains the corpus-specific
instructions used on \textsc{BCP-S}.

\noindent The following listing reproduces the agent-facing manual verbatim except for line
wrapping and the replacement of its single Unicode right arrow with the ASCII digraph
\texttt{->} for typesetting compatibility. Because the listing is verbatim, it retains the
manual's original \emph{document}, \emph{read}, and \emph{open} terminology.

\begin{table}[H]
\centering
\caption{Complete agent-facing Search--Fetch manual, reproduced verbatim in one full-width column.}
\label{tab:agent-manual}
\smallskip\hrule\smallskip
\noindent
\begin{minipage}[t]{\textwidth}
\def\AgentManualFontSize{\fontsize{6pt}{6.4pt}\selectfont}
\begin{Verbatim}
# Structured document search: field-tagged `term[field]` Boolean + fetch

Search the corpus with a Boolean query language: one expression selects the documents whose named fields contain your words. It matches
words, not meaning — no embeddings, so exact terms matter (`OR` or `*` for variants). The language: `term[field]`, `AND`/`OR`/`NOT`,
parentheses, wildcard `*`, quoted `"phrase"`. A **search never shows document bodies** — only structure (title, matched fields, section
names, infobox keys). **fetch** then pulls the one slice that should hold the fact. Terms are case-insensitive. Two moves, in order:

1. **search** a query -> ranked DOCUMENTS: each hit shows the title, the fields your terms matched, the doc's section names, and its infobox
   keys — numbered for `fetch`.
2. **fetch** a named section (or `infobox`) of a numbered doc — a name from that doc's list — to read its text. Never the whole document.

## How to search

1. Query entity NAMES, never the question's wording — relation words (spouse, owner, founder) are what you look FOR in the fetched slice,
   not what you search for. Unsure between spellings? `OR` them: `zurich[title] OR "zürich"[title]`. When the question only DESCRIBES a
   thing (no name given), don't AND the whole description — each added clause loses more docs, so a long AND of common words 0-hits. Start
   from the 1–2 tokens most likely to appear VERBATIM in the target doc — a proper name, a domain term, an exact number like `1897` —
   not the framing words (described, mentioned).
2. Field-scope tightly. `x[title]` = the doc is ABOUT x; `x[body]` = x is merely mentioned somewhere; `x[tiab]` = title-or-body; `x[infobox]`
   = the reverse link — whose FACTS name x (a founder, an owner) when x has no page of its own.
3. Pick the hit by the TYPE the question implies, from the listing, not a fetch. Infobox keys type a hit at a glance — Released/Label = a
   work, Born/Spouse = a person, Country/State = a place. A title carrying a work marker — `(album)`/`(film)`/`(song)` — is that work,
   not a same-named person or place.

## The fields

| field | matches | reach for it when |
|---|---|---|
| `title` \| `ti` | the document's name | the entity should be the doc's SUBJECT |
| `section` \| `sec` | the heading names | a doc devotes a whole section to it |
| `body` \| `ab` \| `text` | the section texts | it's merely mentioned, not the subject |
| `infobox` \| `ib` | the key: value facts | the reverse link — whose facts name it |
| `tiab` | title OR body (combo) | a first broad pass before narrowing |

Not every corpus has real sections/infobox — a flat (non-Wikipedia-structured) document is just title + body, so `[section]`/`[infobox]`
0-hit there on EVERY doc; if one 0-hits immediately, fall back to `[body]`/`[tiab]` rather than re-trying it on a different entity.

Atoms: a bare term matches anywhere in the scoped field. Several bare words parse as one exact adjacent PHRASE — reliably 0-hits unless
truly contiguous (an album title). `word*` widens to any token starting with it. `term[f1,f2]` matches if EITHER field has it. Combine: `A OR
B` (either), `A AND B` (both — use sparingly; 3+ clauses usually over-specify and 0-hit), `A NOT B`. Parentheses group as expected.

## Fetch — reading the slice you found

`search` lists section names as `§[History·Career·Legacy]` and infobox keys as `ib[Born·Spouse]`. `fetch` takes `[rank, section]` pairs
against that numbering:

```
{"specs": [[1, "infobox"]]}
```

- Fetch the ONE slice the fact lives in — the section names already told you where: `infobox` for relational facts, an early section for
  what/who it is.
- If that slice truncates or lacks the fact, fetch a DIFFERENT named section that would hold it (biography / career / legacy / discography),
  not the same intro again — an unopened named section is a lead; never abandon a retrieved doc or fill the gap from memory.
- `section` is a name from that doc's `§[...]` list (or `infobox`), NEVER a doc id, the question's wording, or a name you only hope exists
  — the `rank` already picks the doc.

## Hops — chaining across searches

Each hop is its own search + fetch, not a bigger query. The fetched slice names the next entity — search THAT. No doc of its own? Flip
direction: search it as `[tiab]`/`[infobox]` instead of `[title]` — the fact usually sits on the page that mentions it. 0 hits means your
SURFACE is wrong, not that the doc is absent — recover in ONE move: drop a long name to its 1–2 most distinctive words, try `word*`, or
move `[title]` to `[tiab]`. Two loosenings of the SAME entity both 0-hit? PIVOT to a different entity the question names; never answer from
memory.

Before you stop: check the fact is the ASKED-FOR TYPE, not just the next entity in the chain. Once a slice shows a fact of the right type,
ANSWER — don't re-search to confirm. The answer is the shortest span COPIED VERBATIM from the slice — exact spelling and accents, and
just the asked-for unit, not a compound (the state alone, not "City, State").

## Worked examples

- `harbor[title]` — the doc about the place; fetch its `infobox` for a relational fact.
- `festival[title] AND film[body]` — which same-named hit is the film (a work), not the event.
- `studio[body]` — nothing is titled for it; find the pages that mention it (reverse link).
- `munoz[title] OR "muñoz"[title]` — accent/spelling variants in one search.
- fetch `{"specs": [[1, "infobox"]]}` — read result 1's facts, then chain or answer.

## Common mistakes

- A comma or "and" meant loosely — write real `AND`/`OR`; `AND` requires ALL clauses.
- `A AND B` to CONNECT two entities from different hops (a person `AND` their team's founder) — they never share ONE document, so it
  0-hits. Hop instead: search A, fetch, then search what you found. A tight `AND` is for ONE entity's own distinctive words (`telescope AND
  1893`).
- Searching the question's framing words (described, mentioned) instead of an entity NAME.
- Fetching section after section — the listing already named which slice to open.
- Re-searching to confirm a fact a slice already shows, or hunting a title equal to the answer — just answer from the slice you already
  have.
- Answering the intermediate entity, or a compound, instead of the TYPE actually asked for.
- Swapping in a more familiar entity that merely sounds like the question's name — search the LITERAL name bare first; only widen if that
  0-hits.
\end{Verbatim}
\end{minipage}
\smallskip\hrule
\end{table}
\subsubsection{Condition-Specific Instructions}
\label{app:skills-diffs}

\paragraph{\textsc{Sieve} on \textsc{BCP-S}.}
The BrowseComp Search--Fetch manual retains the same two-step
workflow and query guidance as the Wikipedia manual but substitutes corpus-specific metadata.
It adds \texttt{author} and \texttt{date} guidance and advises falling back to \texttt{body} when
an author is unavailable.

\paragraph{Appendix-only Indri comparison.}
Indri uses a distinct query language and therefore receives its own manual. It describes Indri's
proximity, synonym, weighted-combination, field, and
date operators rather than BQL syntax. Unlike the BQL instructions, it is not customized by
collection. This comparison is included only to situate BQL against an established structured
retrieval system.

\subsection{BQL Queries and Worked Interaction}
\label{app:bql-examples}

To make the workflow concrete, this appendix reproduces a sample of issued \textbf{BQL} queries
and one complete \textsc{Sieve} interaction. The query strings and observations come from experiment
logs for \textbf{Tongyi-DeepResearch-30B-A3B} on
\textsc{BCP-S}, \textsc{HotpotQA-Structured} (HotpotQA-S), and
\textsc{MuSiQue-Structured} (MuSiQue-S). Appendix~\ref{app:grammar} gives the full grammar.
Instance identifiers abbreviate their dataset prefix (e.g.\
\texttt{browsecomp\_plus\_structured\_\_772} is shown as \texttt{\_\_772}).

\subsubsection{Sample of Issued Queries}
\label{app:bql-examples-queries}

Table~\ref{tab:bql-query-sample} lists 12 queries the agent actually issued, chosen to cover the
grammar's main constructs. \emph{Hits} is the number of webpages the query matched; ``0
(fallback)'' marks a query with zero exact Boolean matches, for which the executor returned a soft
BM25-ranked list and advised the agent to verify, loosen, or replace the query
(\S\ref{sec:method-search}). One query is shortened for space, with the elision marked
``\dots{}''; every other query is reproduced exactly as issued.

\begin{table*}[t]
\centering
\caption{Examples of BQL expressions issued by \textsc{Sieve}. \emph{Instance} identifies the
collection and question; \emph{Hits} is the number of exact matches. ``0 (fallback)'' indicates
that no exact match was found and the ranked fallback was returned. ``ti'' and ``tiab'' denote
title and title-or-body fields, respectively.}
\label{tab:bql-query-sample}
\scriptsize
\setlength{\tabcolsep}{4pt}
\begin{tabularx}{\textwidth}{@{}
  >{\raggedright\arraybackslash}p{0.14\textwidth}
  >{\raggedright\arraybackslash}X
  >{\raggedright\arraybackslash}p{0.23\textwidth}
  >{\raggedleft\arraybackslash}p{0.10\textwidth}
@{}}
\toprule
Construct & Query (verbatim) & Instance & Hits \\
\midrule
Plain term & \texttt{EU} &
  MuSiQue-S \texttt{4hop2\_\_9988\allowbreak\_158279\allowbreak\_70784\allowbreak\_79935} & 277 \\
Quoted phrase & \texttt{"Night Shade Books"} & BCP-S \texttt{\_\_870} & 1 \\
Field-scoped & \texttt{Club[ti]} &
  HotpotQA-S \texttt{5ab52996\allowbreak 55429905\allowbreak 94ba9d1e} & 133 \\
Implicit AND (unquoted natural language) &
  \texttt{township established in the 1960s to accommodate migrant workers streets were named
  \dots{} renamed late 2010s} &
  BCP-S \texttt{\_\_772} & 1 \\
\midrule
Boolean AND & \texttt{FDA AND antibiotic} &
  MuSiQue-S \texttt{2hop\_\_35686\allowbreak\_58556} & 3 \\
Boolean OR & \texttt{"Hampshire" OR "Hampden"} &
  HotpotQA-S \texttt{5a838944\allowbreak 55429964\allowbreak 88c2e450} & 991 \\
Negation (NOT) & \texttt{Splendor NOT American[tiab]} &
  HotpotQA-S \texttt{5a77ca39\allowbreak 55429967\allowbreak ab1052a3} & 33 \\
\midrule
Wildcard & \texttt{WNP*} &
  HotpotQA-S \texttt{5a8dcbc0\allowbreak 55429906\allowbreak 8b959df4} & 4 \\
Date range & \texttt{"glassmaker"[body] OR "stained glass artist"[body] AND "died 1880"[date]} &
  BCP-S \texttt{\_\_802} & 30 \\
Multi-constraint &
  \texttt{"Harry Potter"[title] OR "Harry Potter"[body] AND "graffiti"[body] AND
  date[2011-01-01]} & BCP-S \texttt{\_\_790} & 24 \\
\midrule
Zero-exact $\to$ fallback (negation) &
  \texttt{Iraq[title] AND NOT "Iraq War"[title]} &
  MuSiQue-S \texttt{4hop2\_\_9988\allowbreak\_261673\allowbreak\_70784\allowbreak\_61381} & 0 (fallback) \\
Zero-exact $\to$ fallback (date range) &
  \texttt{Kwesi Arthur debut album[title] AND 2018..2023[date]} & BCP-S \texttt{\_\_787} & 0 (fallback) \\
\bottomrule
\end{tabularx}
\end{table*}

\subsubsection{A Worked Interaction}
\label{app:bql-examples-interaction}

We trace \texttt{browsecomp\_plus\_structured\_\_500} in full: a correct three-step interaction
that shows the complete search~$\to$~inspect~$\to$~fetch cycle. The question is:

\begin{quote}
\footnotesize
Identify the English name of structure originally constructed during the 16th century. It was
located in a capital city of a country that was under foreign rule for centuries. The structure
faced destruction three years after its initial completion due to an attack led by a country on
the same continent. Rebuilt within the same century it was initially constructed, it was
eventually damaged again in the mid 17th century.
\end{quote}

\textbf{Step 1 (search).} The agent issues
\texttt{"Fort Santiago"[title] OR "Fort Santiago"[body]}, which compiles to
\texttt{IN(title, "Fort Santiago") OR IN(body, "Fort Santiago")} and returns 5 matches. The
top result card shows (abridged):

{\footnotesize\begin{quote}
\texttt{1  42282  'Fort Santiago - Wikipedia'  \S{}[Naming and role$\cdot$Origins
and construction$\cdot$British occupation\dots{}]  matched: title,body \dots{}}
\end{quote}}

\noindent with four further cards for webpages 86640, 24089, 18850, and 35629 (a Manila city
webpage).

\textbf{Step 2 (fetch).} After inspecting the section headings on the result cards, the agent
requests \texttt{fetch specs = [[1,"Origins and construction"],[3,"(intro)"]]}. This
returns the actual section text, including: ``\emph{The fort was destroyed in 1574 during the
Chinese attack led by Limahong. The stone fort was built between 1589 and 1592 and was repaired
and extended after being damaged by the 1645 earthquake}'' (from webpage 24089's introduction),
and a longer passage from webpage 42282's ``Origins and construction'' section describing the
1571 founding, the 1574 Limahong raid, and the 1590--1593 stone rebuild.

\textbf{Step 3 (answer).} Before answering, the model checks all four constraints against the
fetched text: 1571 construction (16th century); Manila, capital of the Philippines under
Spanish rule for centuries; the 1574 Limahong attack, three years after 1571; the 1589--1592
stone rebuild in the same century; and the 1645 earthquake damage, in the mid-17th century. It
then emits \texttt{<answer>Fort Santiago</answer>}. The gold answer is \emph{Fort Santiago};
the answer is scored exact-match correct.

\section{Experimental Details and Baselines}
\label{app:evaluation-baselines}

This appendix follows the experimental setup in the main paper: baseline implementation details,
common serving and budget settings, evaluation procedures, and retrieval-engine validation.

\subsection{DCI Reimplementation}
\label{app:dci}

Our \textbf{DCI} condition adapts \citet{dci2026} to our evaluation harness
(\S\ref{sec:baseline-systems}). The agent receives only \texttt{bash} and \texttt{read} tools and
searches the raw corpus directly, without a retriever. Where the paper leaves an implementation
detail unspecified, we follow its stated intent. Here, \texttt{read} is the literal name of DCI's
file-access tool. Two deliberate differences remain.

First, we omit the original five-step ``SEARCH STRATEGY (follow exactly)'' prompt. This makes DCI
consistent with our other conditions that do not expose a query language, but likely disadvantages
it relative to the published implementation. Query-language conditions do receive a tool manual,
so prompt support is not fully matched across the experiment (Appendix~\ref{app:fairness}).

Second, our DCI agent receives at most $100$ steps on every collection
(Appendix~\ref{app:setup}). The original paper uses $300$ turns for its main results and reports
that DCI needs more tool calls than a retrieval agent to locate an initial anchor webpage. Our
smaller budget may therefore understate DCI's performance.

The \textbf{BM25-bounded DCI} control follows the central RISE design
\citep{rise2026}: BM25 first retrieves ten webpages for the original question, after which
\texttt{bash} and \texttt{read} operate only on a flat-file staging area containing those
webpages. It uses the same BM25 implementation as the Search--Visit baseline and the same shell
tools as DCI. The bound is enforced by the staged filesystem rather than by prompt instruction.

\subsection{Decoding, Serving, and Budgets}
\label{app:setup}

All agent backbones use temperature $0.6$, seed $42$, local vLLM serving
\citep{kwon2023vllm}, and the released ReAct scaffold \citep{yao2023react}. The loop format, system
prompt, and termination contract are fixed across conditions. The content-access setup is
not the only varying factor, however; Appendix~\ref{app:fairness} describes the remaining
asymmetries.

\paragraph{Context window and serving concurrency.} The primary backbone uses a
$131{,}072$-token context window, set by vLLM's \texttt{--max-model-len}. Qwen-AgentWorld
and OpenResearcher use $262{,}144$ tokens. Each backbone keeps the same context window
when comparing Search--Visit with \textsc{Sieve}. We introduced an early-stop rule part-way through
the project. Once a run reaches $90\%$ of its context window, the rule asks for a final answer instead of waiting for the
step cap to overflow the window. Later cells use this rule, whereas earlier cells do not. Each
run's configuration file records whether it was active.

Serving concurrency ranges from two to eight vLLM workers. We use fewer workers for
long-context datasets to prevent throughput degradation. Because runs are independent,
concurrency changes scheduling rather than generation or scoring. Context early stopping and
concurrency are not matched by construction; the released analysis records any differences within
a reported pair.

\paragraph{Step budgets.}
Every search call returns at most five result cards ($k{=}5$), and every iterative condition has a
maximum budget of $100$ agent steps. These values are fixed in the run configuration rather
than inferred from observed trajectories; runs may stop earlier after producing a final answer
or reaching the context-window threshold. Single-call retrieve-then-read conditions have no
iterative step budget.

\paragraph{Budget comparability.}
Because the cap is uniform, the reported comparisons are not confounded by different nominal step
budgets. They can still differ in realized trajectory length: an agent may answer early, trigger
the context-window stop, or use the complete budget.

\subsection{Evaluation and Answer Recovery}
\label{app:fairness}

\paragraph{Recovering missing final answers.}
Some agents reach the step limit while still using tools and never produce a scorable final answer.
Because the frequency of these empty runs differs across conditions, we separately evaluate
whether a terminal answer can be recovered from the completed trajectory.

We apply a deterministic offline recovery procedure to empty runs. It replays the terminal
transcript with the assistant response prefilled by \texttt{<answer>}, which permits only an
answer continuation and no further tool use. The procedure recovers about $93\%$ of the runs to
which it is applied. It writes a separate overlay rather than modifying the original rows, and it
runs before judged comparisons because recovered answers can change system rankings.

Recovery coverage is unfortunately incomplete. All \textsc{BCP-S} cells and
the two Wikipedia Search--Visit baselines have recovery overlays. The Wikipedia \textsc{Sieve}
arms and other Search--Fetch conditions do not. This asymmetry limits
comparisons across those groups.

\paragraph{Matching token budgets.}
For retrieve-then-read baselines, we enforce the budget after assembling and tokenizing the complete
prompt, including the chat template. A fit loop removes content until the prompt fits the model
context window.

For agentic conditions, whole-webpage visit and section fetch share a
$12{,}000$-token ceiling per webpage. Their comparison therefore changes what content is returned, not the
maximum amount available from a webpage. All iterative conditions also use the same maximum budget
of $100$ agent steps, and every search call returns at most five results. A post-hoc audit
identified two remaining asymmetries: tool documentation and result-list rendering. We give the
direction of each below and reproduce the condition-specific instructions in
Appendix~\ref{app:skills}.

\paragraph{Validating engine reimplementations.}
We compare reference retrieval components with their production engines on identical inputs before
interpreting an experimental difference. Appendix~\ref{app:engine} reports the divergence we found
rather than treating the implementations as equivalent.

\subsection{Retrieval Engine Controls}
\label{app:engine}

Zero-hit recovery has two stages. For a multi-constraint query, \textsc{Sieve} first returns partial
matches, ranked by the number of satisfied constraints and then by the condition's configured
ranker within each coverage tier. If no partial match exists, it ranks the full collection using a
dependency-free BM25 approximation over the positive query terms. A zero-hit single-constraint
query uses this second stage directly.

On identical queries and corpora, the approximation's top-5 Jaccard overlap with Lucene BM25 is
$0.546$ on \textsc{BCP-S}, indicating substantial ranking divergence. Zero-hit
fallbacks account for $36.8$--$53.7\%$ of \textsc{Sieve}'s search calls. The reported results
therefore use Lucene for Boolean filtering and ordinary sparse ranking, with the approximate BM25
scorer used within the fallback path; they should not be interpreted as a pure Lucene evaluation
(\S\ref{sec:models-retrieval}).

\paragraph{Comparison with an established structured executor.}
We additionally report Indri \citep{strohman2005indri} to situate BQL against an established
system for Boolean and structured queries. Ordered and unordered proximity operators compile to
Lucene \texttt{SpanNearQuery}; typed dates compile to fielded range queries; and belief-combination
operators use \texttt{LMDirichletSimilarity} with $\mu{=}2500$ \citep{zhai2001study}. This
condition is excluded from the main system comparison.

Table~\ref{tab:sieve-indri} compares \textsc{Sieve} with an Indri-style executor while holding
result cards, section-fetch behavior, result-pool size, and content-access budget fixed. This comparison
places the proposed executor against an established structured-retrieval alternative; it is not
part of the comprehensive system comparison in Table~\ref{tab:consolidated}.

\begin{table*}[t]
\centering
\caption{Structured-retrieval executor comparison on \textsc{BCP-S}. Judge and EM report answer
accuracy, recall measures gold-webpage exposure, and the remaining columns report context use and
model calls. Bold marks the best value in each column; asterisks denote significance versus
\textsc{Sieve} after Bonferroni correction.}
\label{tab:sieve-indri}
\small
\setlength{\tabcolsep}{5pt}
\begin{tabular*}{\textwidth}{@{\extracolsep{\fill}}lrrrrrr@{}}
\toprule
Condition & Judge\% & EM\% & Recall\% & Tok./inst & Acc.\ tok./inst & Calls \\
\midrule
Indri & 30.6* & 28.0 & 46.9 & 47k & 1660k & 69.5* \\
Indri + dense belief & 32.3* & 30.1 & 48.9 & 48k & 1671k & 68.5* \\
\textsc{Sieve} & \textbf{37.2} & \textbf{34.5} & \textbf{56.6} & \textbf{46k} & \textbf{1593k} & \textbf{62.2} \\
\bottomrule
\end{tabular*}
\end{table*}

\section{Full Results and Additional Analyses}
\label{app:additional-results}

This appendix follows the order of the main findings: complete system results, the broader
efficiency--effectiveness comparison, component and robustness analyses, and failure diagnostics.

\subsection{Complete System Results}
\label{app:fulltables}

The following tables expand each collection column in Table~\ref{tab:consolidated}. \emph{Recall}
records whether a gold webpage was surfaced; for one-shot systems, it instead uses the retrieved
set. \emph{Tok./inst} counts distinct tokens, \emph{Acc.\ tok./inst} sums the growing prompt across
calls, and \emph{Calls} reports the number of model calls. The tables follow the main-text order:
HotpotQA, MuSiQue, and \textsc{BCP-S}.


\begin{table*}[t]
\centering
\caption{Full HotpotQA results for the systems in Table~\ref{tab:consolidated}. Superscripts and
bolding follow Table~\ref{tab:consolidated}.}
\label{tab:hotpotqa-results}
\footnotesize
\begin{tabular*}{\textwidth}{@{\extracolsep{\fill}}>{\raggedright\arraybackslash}p{5.6cm}rrrrr@{}}
\toprule
Condition & EM\% & Recall\% & Tok/inst & Acc.\ tok/inst & Calls \\
\midrule
\multicolumn{6}{l}{\emph{One-shot retrieve-then-read}} \\
BM25 & 29.9$^{ab}$ & 86.9 & 1k$^{ab}$ & 28k & 1.0$^{ab}$ \\
Dense & 30.3$^{ab}$ & 90.1 & 1k$^{ab}$ & 15k & 1.0$^{ab}$ \\
\midrule
\multicolumn{6}{l}{\emph{Iterative Search--AutoRead}} \\
BM25 & 42.4$^{b}$ & 91.0 & 72k$^{ab}$ & 3226k & 36.8$^{ab}$ \\
Dense & 39.9$^{ab}$ & 85.8 & 62k$^{ab}$ & 2980k & 35.8$^{ab}$ \\
\midrule
\multicolumn{6}{l}{\emph{Direct corpus interaction}} \\
DCI (no retriever) & 42.8$^{b}$ & 94.3 & 35k$^{ab}$ & 861k & 21.8$^{a}$ \\
BM25-bounded DCI (RISE-style) & 42.6$^{b}$ & 95.3 & 21k$^{ab}$ & 431k & 22.1$^{ab}$ \\
\midrule
\multicolumn{6}{l}{\emph{Iterative Search--Visit (whole-webpage visits)}} \\
BM25 & 43.7$^{b}$ & 95.2 & 20k$^{b}$ & 295k & \textbf{15.9}$^{b}$ \\
Dense & 43.0$^{b}$ & 93.3 & 21k$^{ab}$ & 563k & 25.9$^{ab}$ \\
BM25+Dense & 40.2$^{ab}$ & 85.9 & 21k$^{ab}$ & 643k & 35.4$^{ab}$ \\
\midrule
\multicolumn{6}{l}{\emph{Iterative Search--Fetch (section fetching)}} \\
BM25 & 41.0$^{ab}$ & 94.9 & \textbf{13k}$^{ab}$ & 350k & 31.1$^{ab}$ \\
Dense & 40.7$^{ab}$ & 93.5 & 14k$^{a}$ & 381k & 33.2$^{ab}$ \\
BM25+Dense & 40.5$^{ab}$ & 95.4 & 14k$^{a}$ & 370k & 32.6$^{ab}$ \\
\cmidrule(lr){1-6}
\textsc{Sieve} (BQL-filtered BM25) & 44.5 & 95.2 & 15k$^{ab}$ & 389k & 27.9$^{ab}$ \\
\textsc{Sieve} (BQL-filtered Dense) & \textbf{45.8}$^{a}$ & \textbf{95.5} & 15k$^{ab}$ & 394k & 27.9$^{ab}$ \\
\textsc{Sieve} (BQL-filtered BM25+Dense) & 45.3$^{a}$ & 94.8 & 14k$^{a}$ & \textbf{224k} & 20.9$^{a}$ \\
\bottomrule
\end{tabular*}
\end{table*}

\begin{table*}[t]
\centering
\caption{Full MuSiQue results for the systems in Table~\ref{tab:consolidated}. Notation follows
Table~\ref{tab:hotpotqa-results}.}
\label{tab:musique-results}
\footnotesize
\begin{tabular*}{\textwidth}{@{\extracolsep{\fill}}>{\raggedright\arraybackslash}p{5.6cm}rrrrr@{}}
\toprule
Condition & EM\% & Recall\% & Tok/inst & Acc.\ tok/inst & Calls \\
\midrule
\multicolumn{6}{l}{\emph{One-shot retrieve-then-read}} \\
BM25 & 7.8$^{ab}$ & 66.0 & 2k$^{ab}$ & 54k & 1.0$^{ab}$ \\
Dense & 11.0$^{ab}$ & 82.6 & 1k$^{ab}$ & 32k & 1.0$^{ab}$ \\
\midrule
\multicolumn{6}{l}{\emph{Iterative Search--AutoRead}} \\
BM25 & 20.9$^{ab}$ & 86.7 & 108k$^{ab}$ & 6508k & 78.3$^{ab}$ \\
Dense & 20.5$^{ab}$ & 79.4 & 95k$^{ab}$ & 5593k & 71.2$^{ab}$ \\
\midrule
\multicolumn{6}{l}{\emph{Direct corpus interaction}} \\
DCI (no retriever) & 26.9 & 93.2 & 63k$^{ab}$ & 2071k & 39.4$^{ab}$ \\
BM25-bounded DCI (RISE-style) & 26.3$^{b}$ & 95.1 & 47k$^{ab}$ & 1328k & 39.6$^{ab}$ \\
\midrule
\multicolumn{6}{l}{\emph{Iterative Search--Visit (whole-webpage visits)}} \\
BM25 & 26.1$^{b}$ & 95.2 & 43k$^{b}$ & 816k & \textbf{28.7}$^{b}$ \\
Dense & 25.5$^{b}$ & 93.3 & 42k$^{b}$ & 1436k & 47.6$^{ab}$ \\
BM25+Dense & 21.8$^{ab}$ & 84.4 & 39k$^{ab}$ & 1363k & 56.3$^{ab}$ \\
\midrule
\multicolumn{6}{l}{\emph{Iterative Search--Fetch (section fetching)}} \\
BM25 & 24.5$^{b}$ & 94.6 & 22k$^{ab}$ & 725k & 50.3$^{ab}$ \\
Dense & 23.4$^{ab}$ & 93.4 & 23k$^{ab}$ & 770k & 53.9$^{ab}$ \\
BM25+Dense & 24.6$^{b}$ & 95.7 & 23k$^{ab}$ & 764k & 53.0$^{ab}$ \\
\cmidrule(lr){1-6}
\textsc{Sieve} (BQL-filtered BM25) & 27.9 & 95.6 & 24k$^{ab}$ & 789k & 47.6$^{ab}$ \\
\textsc{Sieve} (BQL-filtered Dense) & 28.2 & \textbf{95.8} & 24k$^{ab}$ & 809k & 48.0$^{ab}$ \\
\textsc{Sieve} (BQL-filtered BM25+Dense) & \textbf{29.2}$^{a}$ & 95.4 & \textbf{21k}$^{a}$ & \textbf{428k} & 32.6$^{a}$ \\
\bottomrule
\end{tabular*}
\end{table*}


\begin{table*}[t]
\centering
\caption{Full \textsc{BCP-S} results for the systems in Table~\ref{tab:consolidated}. Notation
follows Table~\ref{tab:consolidated}.}
\label{tab:bcp-full}
\small
\begin{tabular*}{\textwidth}{@{\extracolsep{\fill}}>{\raggedright\arraybackslash}p{5.6cm}rrrrrr@{}}
\toprule
Condition & Judge\% & EM\% & Recall\% & Tok/inst & Acc.\ tok/inst & Calls \\
\midrule
\multicolumn{7}{l}{\emph{One-shot retrieve-then-read}} \\
BM25 & 2.2$^{ab}$ & 1.1 & 3.0 & 2k$^{ab}$ & 119k & 1.0$^{ab}$ \\
Dense & 6.9$^{ab}$ & 5.5 & 9.0 & 2k$^{ab}$ & 21k & 1.0$^{ab}$ \\
\midrule
\multicolumn{7}{l}{\emph{Iterative Search--AutoRead}} \\
BM25 & 13.6$^{ab}$ & 12.2 & 20.0 & 142k$^{ab}$ & 6716k & 85.4$^{ab}$ \\
Dense & 13.0$^{ab}$ & 12.3 & 24.1 & 122k$^{ab}$ & 6179k & 76.3$^{ab}$ \\
\midrule
\multicolumn{7}{l}{\emph{Direct corpus interaction}} \\
DCI (no retriever) & 22.2$^{ab}$ & 20.2 & 40.2 & 107k$^{ab}$ & 3453k & \textbf{53.6}$^{ab}$ \\
BM25-bounded DCI (RISE-style) & 33.4 & 30.6 & 55.2 & 75k$^{ab}$ & 2640k & 64.0 \\
\midrule
\multicolumn{7}{l}{\emph{Iterative Search--Visit (whole-webpage visits)}} \\
BM25 & 34.7 & 31.1 & 53.1 & 68k$^{b}$ & 2387k & 64.1 \\
Dense & 36.5 & 33.9 & 57.7 & 58k$^{ab}$ & 1915k & 65.7 \\
BM25+Dense & 33.7 & 30.8 & 52.3 & 53k$^{ab}$ & 1802k & 68.5$^{b}$ \\
\midrule
\multicolumn{7}{l}{\emph{Iterative Search--Fetch (section fetching)}} \\
BM25 & 30.6$^{b}$ & 28.2 & 52.9 & 40k$^{ab}$ & 1450k & 76.6$^{ab}$ \\
Dense & 30.1$^{b}$ & 27.8 & 52.9 & 37k$^{ab}$ & 1362k & 78.6$^{ab}$ \\
BM25+Dense & 34.9 & 32.7 & \textbf{62.9} & \textbf{35k}$^{ab}$ & \textbf{1234k} & 73.6$^{ab}$ \\
\cmidrule(lr){1-7}
\textsc{Sieve} (BQL-filtered BM25) & 33.6 & 31.7 & 53.0 & 49k$^{a}$ & 1643k & 62.0 \\
\textsc{Sieve} (BQL-filtered Dense) & 34.6 & 32.5 & 52.0 & 48k$^{a}$ & 1667k & 61.5 \\
\textsc{Sieve} (BQL-filtered BM25+Dense) & \textbf{37.2} & \textbf{34.5} & 56.6 & 46k$^{a}$ & 1593k & 62.2 \\
\bottomrule
\end{tabular*}
\end{table*}

\clearpage

\subsection{Efficiency--Effectiveness Comparison}
\label{app:efficiency}

\begin{figure}[H]
\centering
\includegraphics[width=0.6\linewidth]{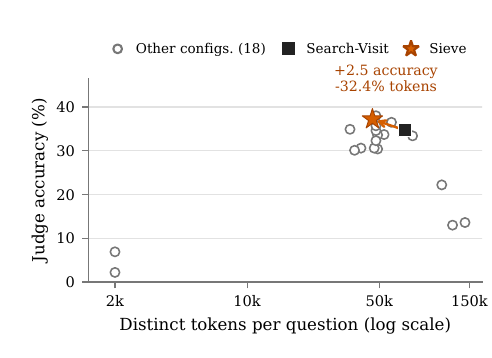}
\caption{Judge accuracy versus distinct tokens for the primary-backbone \textsc{BCP-S}
configurations. Open circles denote other configurations, the square marks BM25 Search--Visit,
and the star marks \textsc{Sieve}; the arrow connects the latter two.}
\label{fig:pareto}
\end{figure}

Figure~\ref{fig:pareto} places all primary-backbone \textsc{BCP-S} configurations on a common
accuracy--context plane, including the retriever-sensitivity and structured-executor controls
reported outside the comprehensive main table.

\subsection{Component and Robustness Analyses}
\label{app:controls}

\paragraph{Query-focused snippets.}
Table~\ref{tab:snippet-ablation} gives the full accuracy and cost
measurements for the matched snippet comparison summarized in Figure~\ref{fig:snippet-ablation}.
The only difference is whether the result card includes the $25$-token query-focused snippet.

\begin{table}[H]
\centering
\footnotesize
\setlength{\tabcolsep}{3.2pt}
\renewcommand{\arraystretch}{0.94}
\caption{Matched snippet ablation. \emph{Acc.} is the judge verdict on \textsc{BCP-S} and exact
match on HotpotQA and MuSiQue; cost columns follow \S\ref{sec:evaluation-protocol}. Bold marks the
better value. Asterisks denote significance versus \textsc{Sieve} after Bonferroni correction.}
\label{tab:snippet-ablation}
\begin{tabular*}{\columnwidth}{@{\extracolsep{\fill}}lrrrr@{}}
\toprule
Condition & Acc. & Tok. & Acc.\ tok. & Calls \\
\midrule
\multicolumn{5}{l}{\emph{HotpotQA}} \\
\textsc{Sieve} without snippets & 42.4* & \textbf{13k}* & 228k & 25.0* \\
\textsc{Sieve} & \textbf{45.3} & 14k & \textbf{224k} & \textbf{20.9} \\
\midrule
\multicolumn{5}{l}{\emph{MuSiQue}} \\
\textsc{Sieve} without snippets & 24.9* & \textbf{19k}* & \textbf{391k} & 35.7* \\
\textsc{Sieve} & \textbf{29.2} & 21k & 428k & \textbf{32.6} \\
\midrule
\multicolumn{5}{l}{\textsc{BCP-S}} \\
\textsc{Sieve} without snippets & 30.4* & 49k & \textbf{1592k} & 66.1* \\
\textsc{Sieve} & \textbf{37.2} & \textbf{46k} & 1593k & \textbf{62.2} \\
\bottomrule
\end{tabular*}
\end{table}

\paragraph{Retriever sensitivity.}
Table~\ref{tab:retriever-sensitivity} gives the complete accuracy and cost measurements summarized
in Figure~\ref{fig:retriever-sensitivity}. Only the dense encoder changes across rows.

\begin{table}[H]
\centering
\caption{Dense-encoder sensitivity on \textsc{BCP-S}. ``Default'' marks the main encoder.}
\label{tab:retriever-sensitivity}
\small
\setlength{\tabcolsep}{3pt}
\begin{tabular*}{\columnwidth}{@{\extracolsep{\fill}}lrrr@{}}
\toprule
Dense encoder & Acc. & Tok. & Calls \\
\midrule
bge-small-en-v1.5 (33M) & 37.2 & 47k & 62.7 \\
bge-base-en-v1.5 (110M, default) & 37.2 & \textbf{46k} & \textbf{62.2} \\
bge-large-en-v1.5 (335M) & 34.6 & 48k & 63.3 \\
Qwen3-Embedding-0.6B & 35.7 & 48k & 65.8 \\
Qwen3-Embedding-4B & \textbf{38.0} & 48k & 64.4 \\
Qwen3-Embedding-8B & 37.0 & 48k & 64.7 \\
\bottomrule
\end{tabular*}
\end{table}

\paragraph{Agent-backbone transfer.}
Table~\ref{tab:backbone-transfer} gives the complete comparison summarized in
\S\ref{sec:ablation-components}. For each backbone and collection, it pairs BM25 Search--Visit with
\textsc{Sieve} under the same serving configuration.

\begin{table}[H]
\centering
\caption{Agent-backbone transfer. Each cell shows BM25 Search--Visit $\rightarrow$ \textsc{Sieve}.
Accuracy is the judge verdict on \textsc{BCP-S} and exact match elsewhere; tokens are distinct
tokens per question. Full model identifiers appear in \S\ref{sec:models-retrieval}.}
\label{tab:backbone-transfer}
\footnotesize
\setlength{\tabcolsep}{2.6pt}
\renewcommand{\arraystretch}{1.0}
\begin{tabular*}{\columnwidth}{@{\extracolsep{\fill}}llrr@{}}
\toprule
Backbone & Collection & Accuracy & Tokens \\
\midrule
\multirow{3}{*}{Tongyi-DeepResearch} & BCP-S & $34.7{\rightarrow}37.2$ & $68.1{\rightarrow}46.0$k \\
& HotpotQA & $43.7{\rightarrow}45.3$ & $19.9{\rightarrow}13.9$k \\
& MuSiQue & $26.1{\rightarrow}29.2$ & $43.0{\rightarrow}21.3$k \\
\midrule
\multirow{3}{*}{Qwen-AgentWorld} & BCP-S & $23.3{\rightarrow}41.1$ & $64.7{\rightarrow}53.6$k \\
& HotpotQA & $40.3{\rightarrow}40.1$ & $13.7{\rightarrow}14.0$k \\
& MuSiQue & $25.7{\rightarrow}29.9$ & $27.5{\rightarrow}25.6$k \\
\midrule
\multirow{3}{*}{OpenResearcher} & BCP-S & $18.7{\rightarrow}25.7$ & $88.6{\rightarrow}55.7$k \\
& HotpotQA & $35.6{\rightarrow}35.0$ & $55.8{\rightarrow}25.8$k \\
& MuSiQue & $20.2{\rightarrow}20.8$ & $76.6{\rightarrow}31.8$k \\
\bottomrule
\end{tabular*}
\end{table}

\paragraph{BQL use and additional controls.}
Table~\ref{tab:operator-adoption} reports how often \textsc{Sieve} uses each part of BQL.

\begin{table}[H]
\centering
\caption{Share of questions for which each BQL operator class appears at least once. Rows overlap
because one question may use several classes. \emph{No field/Boolean/date} excludes those three
operator classes but may include phrases or wildcards.}
\label{tab:operator-adoption}
\small
\setlength{\tabcolsep}{4pt}
\begin{tabular*}{\columnwidth}{@{\extracolsep{\fill}}lrrr@{}}
\toprule
Operator class & \textsc{BCP-S} & HotpotQA & MuSiQue \\
\midrule
Quoted phrase      & 88.4 & 76.0 & 89.2 \\
Field restriction  & 42.8 & 66.0 & 69.1 \\
Boolean            & 34.1 & 41.9 & 44.1 \\
Wildcard           & 1.8  & 4.9  & 8.4  \\
Typed date range   & 0.7  & 0.2  & 0.1  \\
Proximity          & 0.0  & 0.0  & 0.0  \\
\midrule
No field/Boolean/date & 54.7 & 30.9 & 30.1 \\
\bottomrule
\end{tabular*}
\end{table}

Table~\ref{tab:additional-controls} tests \textsc{Sieve} without ranked fallback and compares
Search--Visit over structured and flat corpora.

\begin{table}[H]
\centering
\caption{Controls for zero-hit fallback, snippets, and field exposure on \textsc{BCP-S}. The top
block modifies \textsc{Sieve}; the bottom block compares Search--Visit over the structured and flat
twins. Bold marks the best value in each column; asterisks denote significance versus
\textsc{Sieve} after Bonferroni correction.}
\label{tab:additional-controls}
\footnotesize
\setlength{\tabcolsep}{2pt}
\begin{tabular*}{\columnwidth}{@{\extracolsep{\fill}}>{\raggedright\arraybackslash}p{3.45cm}rrrr@{}}
\toprule
Condition & Judge\% & EM\% & Tok. & Calls \\
\midrule
\textsc{Sieve} & \textbf{37.2} & \textbf{34.5} & \textbf{46k} & \textbf{62.2} \\
Strict Boolean (no fallback) & 30.8* & 29.3 & 46k & 73.6* \\
Without snippets & 30.4* & 28.3 & 49k & 66.1* \\
\midrule
Search--Visit, structured corpus & 34.7 & 31.1 & 68k* & 64.1 \\
Search--Visit, flat corpus & 33.1 & 29.8 & 69k* & 64.1 \\
\bottomrule
\end{tabular*}
\end{table}

\subsection{Hand-Traced Failure Examples}
\label{app:traces}

We manually examine five incorrect (EM${=}0$) \textsc{Sieve} runs from
\textsc{BCP-S} to check the boundaries used by the automatic failure
decomposition in Section~\ref{sec:ablation-failures}. These are diagnostic examples rather than a
representative sample.

\paragraph{Retrieval failures.}
Instances \texttt{\_\_775}, \texttt{\_\_770}, and \texttt{\_\_774} have the gold answers
\emph{Boston}, \emph{Laura Lojo-Rodriguez}, and \emph{Georgia Hirst}, respectively; the model
instead answers \emph{Santiago}, \emph{Muhammad Faruque}, and \emph{1845}. No gold webpage
identifier appears in any search observation. These runs are therefore labeled retrieval
failures irrespective of which result the agent later chooses or how it reasons over that result.

\paragraph{Selection failure.}
For \texttt{\_\_788}, the gold answer is \emph{Taj-ul-Masajid}, while the model returns
\emph{Faisal Mosque}. The result list contains three of the four gold webpage identifiers, so the
necessary evidence is surfaced. Both subsequent fetches, however, target sections from a
``largest mosque list'' webpage rather than any of those gold webpages. The evidence is retrieved but
never fetched, matching the selection-failure definition.

\paragraph{Synthesis failure.}
For \texttt{\_\_773}, the gold answer is \emph{Red}, but the model answers \emph{white}. Here the
agent fetches a section from a gold webpage containing the answer before responding. Retrieval and selection have
therefore succeeded; the remaining error lies in extracting or reasoning to the final answer.

These traces support Figure~\ref{fig:failure-decomposition}'s distinction among retrieval,
selection, and synthesis failures.

\end{document}